\documentclass[aps,prl,amsmath,10pt,superscriptaddress,floatfix,nofootinbib]{revtex4-1}

\usepackage{graphicx}
\usepackage{dcolumn}
\usepackage{bm}
\usepackage{multirow}
\usepackage{overpic}
\usepackage{bm}

\usepackage[colorlinks=true,linktocpage=true,linkcolor=blue,citecolor=blue]{hyperref}

\begin{document}

\title{Study of exotic hadrons with machine learning}

\author{Jiahao Liu}
\address{Guangdong Provincial Key Laboratory of Nuclear Science, Institute of Quantum Matter, South China Normal University, Guangzhou 510006, China}
\affiliation{Guangdong-Hong Kong Joint Laboratory of Quantum Matter, Southern Nuclear Science Computing Center, South China Normal University, Guangzhou 510006, China}

\author {Zhenyu Zhang}
\address{Guangdong Provincial Key Laboratory of Nuclear Science, Institute of Quantum Matter, South China Normal University, Guangzhou 510006, China}
\affiliation{Guangdong-Hong Kong Joint Laboratory of Quantum Matter, Southern Nuclear Science Computing Center, South China Normal University, Guangzhou 510006, China}

\author{Jifeng Hu}\email{hujf@m.scnu.edu.cn}
\address{Guangdong Provincial Key Laboratory of Nuclear Science, Institute of Quantum Matter, South China Normal University, Guangzhou 510006, China}
\affiliation{Guangdong-Hong Kong Joint Laboratory of Quantum Matter, Southern Nuclear Science Computing Center, South China Normal University, Guangzhou 510006, China}

\author{Qian Wang}\email{qianwang@m.scnu.edu.cn}
\address{Guangdong Provincial Key Laboratory of Nuclear Science, Institute of Quantum Matter, South China Normal University, Guangzhou 510006, China}
\affiliation{Guangdong-Hong Kong Joint Laboratory of Quantum Matter, Southern Nuclear Science Computing Center, South China Normal University, Guangzhou 510006, China}
\date{\today}

\begin{abstract}

We analyzed
the invariant mass spectrum of near-threshold exotic states for one-channel candidates with a  deep neural network.
 It can extract the scattering length and effective range,
 which would shed light on the nature of given states,
 from the experimental mass spectrum. As an application,
 the mass spectrum of the $X(3872)$ and the $T_{cc}^+$
 are studied. The obtained scattering lengths, effective ranges, and most relevant thresholds
are consistent with those from fitting to the experimental data.
The advantage of the neural network is that it is more stable than the fitting,
especially for low-statistic data.
The network, which provides another way to analyze the experimental data,
 can also be applied to other one-channel near-threshold exotic candidates.
 \end{abstract}

\maketitle

\section{Introduction}

The color confinement property of quantum chromodynamics (QCD) allows for the existence of any color
neutral object. That challenges the conventional quark model,
in which hadrons are made of either quark-antiquark (mesons) or three quarks (baryons).
Especially, the observed exotic hadrons beyond the conventional configurations
provide a way to decode the mystery of hadronization. Up to now,
tens of exotic candidates have been reported and studied from various aspects
~\cite{Chen:2016qju,Chen:2016spr,Dong:2017gaw,Lebed:2016hpi,Guo:2017jvc,Liu:2019zoy,Albuquerque:2018jkn,Yamaguchi:2019vea,Guo:2019twa,Brambilla:2019esw}.
One important feature is that most of them are very close
to nearby thresholds, as an indication of a large mixture of
continuum~\cite{Guo:2017jvc}. In principle, all the configurations with the same quantum
number can mix with each other. However, which configuration plays an important role,
 either a large size hadronic molecule or compact object is still a well-established question.
 The key value is the probability $1-\lambda^2$
  (with $\lambda^2$ the wave function renormalization constant) of finding continuum in
 a given physical state.  A typical example is deuteron,
 for instance , see Refs.~\cite{Weinberg:1962hj,Weinberg:1965zz} discussed by Weinberg in the 1960s.
 This method has been intensively used
for discussing the nature of exotic candidates in both the hadronic
molecular picture~\cite{Guo:2017jvc} and compact one~\cite{Esposito:2016noz}.
 The value of $\lambda^2$ is related to scattering length~\cite{Guo:2017jvc},
 \begin{eqnarray}
  a&=&-2\frac{1-\lambda^2}{2-\lambda^2}\left(\frac{1}{\gamma}\right)+\mathcal{O}\left(\frac{1}{\beta}\right)
  \label{eq:scatteringlength}
\end{eqnarray}
 and effective range
 \begin{eqnarray}
  r&=&-\frac{\lambda^2}{1-\lambda^2}\left(\frac{1}{\gamma}\right)+\mathcal{O}\left(\frac{1}{\beta}\right)
  \label{eq:effectiverange}
\end{eqnarray}
  of the elastic channel~\cite{Guo:2017jvc} for the one-channel case in the low-energy limit,
  which means that the formulas work in the near-threshold energy region.
  Here, $\gamma=\sqrt{2\mu E_B}$ is the binding momentum with reduced mass $\mu$
  and binding energy $E_B$. $\frac{1}{\beta}$ is the order of range correction.
Here, $\lambda^2=0$ and $\lambda^2=1$ are for the pure molecule and compact object, respectively.
In other words, to the leading order,
\begin{eqnarray}
  a=-\frac{1}{\gamma},\quad r=\mathcal{O}\left(\frac{1}{\beta}\right)
\end{eqnarray}
for the pure molecule and
\begin{eqnarray}
  a=-\mathcal{O}\left(\frac{1}{\beta}\right),\quad r=-\infty
\end{eqnarray}
for the compact object. As a result, extraction of the scattering length and effective range
from the experimental data is a direct way to shed light on the nature of interested hadrons.
Recent and typical examples are the $X(3872)$~\cite{LHCb:2020xds,Baru:2021ldu,Esposito:2021vhu}
and $T_{cc}^+$~\cite{LHCb:2021auc,Baru:2021ldu,Du:2021zzh,Albaladejo:2021vln}
from both experimental and theoretical sides. This work aims at developing
a deep learning network for automatically extracting the scattering length and effective range
from experimental data directly. The final goal is to set up a deep learning network implementing a
multichannel case. As the first step, this work starts from the one-channel case.
This method has been successfully applied to the $P_c(4312)$~\cite{Ng:2021ibr},
the $\pi N$ system~\cite{Sombillo:2021rxv,Sombillo:2021yxe},
 and the nucleon-nucleon system~\cite{Sombillo:2020ccg,Sombillo:2021ifs},
 focusing on various facts.
 For instance, Ref.~\cite{Ng:2021ibr} sets a classifier, instead of extracting
 scattering length and effective range, of a given state by a bottom-up
 approach to avoid model dependence.
 
 \begin{figure}
  \centering
\begin{overpic}[width=8.0cm,height=6.0cm,angle=0]{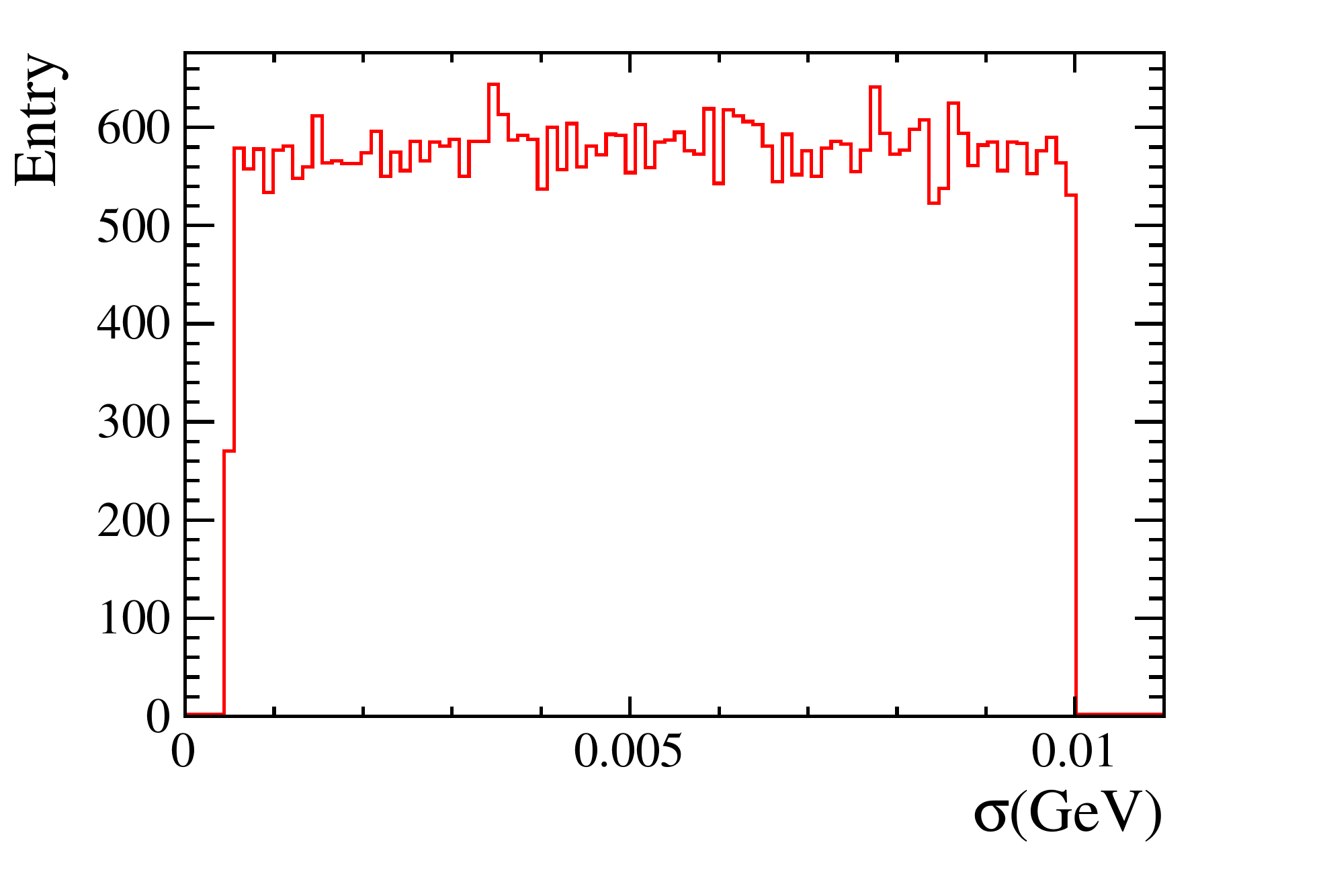}
\put(0,72){$(a)$}
\end{overpic}
\begin{overpic}[width=8.0cm,height=6.0cm,angle=0]{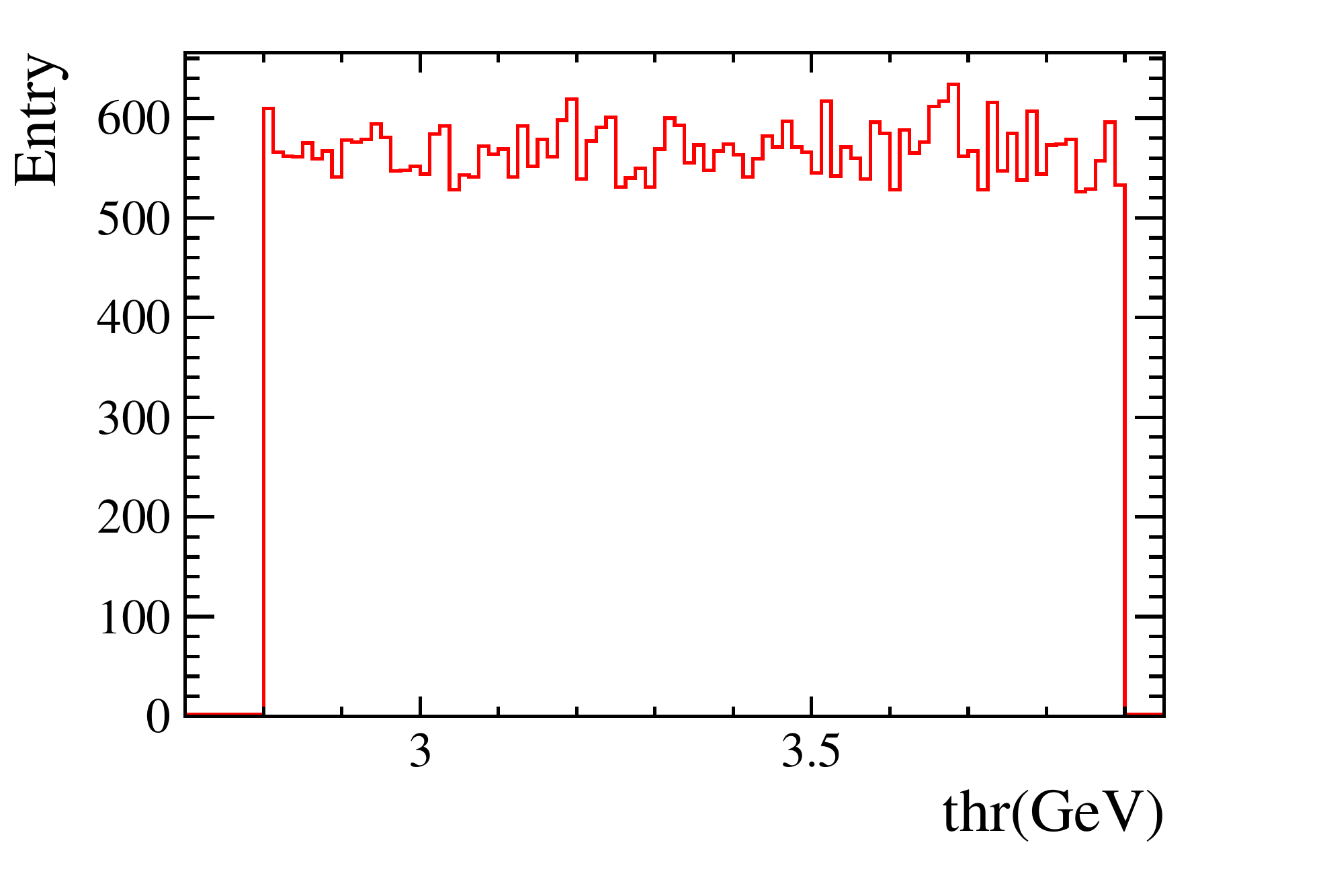}
\put(0,72){$(b)$}
 \end{overpic}
  \caption{ The distributions for the parameter threshold and $\sigma$ at the generation level. }
  \label{parameters_distribution}
  \end{figure}

\section{Physics framework}
The expressions of scattering length [Eq.(\ref{eq:scatteringlength})] and effective range [Eq.(\ref{eq:effectiverange})]
are obtained by matching the effective range expansion scattering amplitude,
\begin{align}
T_{NR}(E)=-\frac{2\pi}{\mu}\frac{1}{1/a+(r/2)k^2-ik}\label{eq:ERE},
\end{align}
to the explicit scattering amplitude, where the subscript ``NR'' indicates
the nonrelativistic expression.
Here, $\mu=\frac{m_1 m_2}{m_1+m_2}$ and $E$ are the reduced mass and total energy of the two-particle system, respectively.
Accordingly, $k=\sqrt{2\mu (E-m_1-m_2)}$ is the three momentum of the scattering particle in the center-of-mass frame.
$m_1$ and $m_2$ are the masses of the two particles.

As the line shape of a state is dominated by the elastic to elastic
\footnote{Here, elastic channel means the channel strongly coupled to the interested state.}
 scattering amplitude~\cite{Dong:2020hxe}
once the elastic channel is predominant in the production vertex,
one can consider that the line shapes are described by $|T_{NR}(E)|^2$ convoluted with a Gaussian function,
\begin{equation}
G(x)=\frac{1}{\sigma\sqrt{2\pi}}e^{-\frac{x^2}{2\sigma^2}},
\label{Gaussian}
\end{equation}
where the mean value is set to zero, up to a phase space factor.
That is,
\begin{align}
PDF(E; a,r, \mathrm{threshold}, \sigma) = \int |T_{NR}(E)|^2 G(E'-E) dE'. \label{eq:pdf}
\end{align}
 The $\sigma$ denotes the resolution which depends on
the energy resolution of measuring the invariant mass spectrum.
Based on the probability density function defined as Eq.~\eqref{eq:pdf}, we generate 150000
line shapes for training with the parameters within the regions
\begin{align}
  a &\in [4.93, 14.80]~\mathrm{fm}, \label{eq:region1}\\
  r &\in [0.49,0.99]\cup[-9.87, -0.49]~\mathrm{fm},\label{eq:region2}\\
  m_1+m_2 &\in [2.8, 3.9]~\mathrm{GeV},\label{eq:region3}\\
  \sigma &\in [0.5, 10]~\mathrm{MeV}. \label{eq:region4}
  \end{align}
  The regions of scattering length and effective range
  allow for both bound and virtual states~\cite{Matuschek:2020gqe}.
  The threshold region covers a charmonium(like) energy region,
as mass resolution actually depends on the momentum resolution,
which is not a constant generally, in the experiment.  The constant resolution is
only an average effect to determine which is a tough job in the experiment.
Thus, we set it as a free parameter to allow for the possibility to extract the resolution from a well-established line shape.
As a result, the resolution region is set to cover the usual experimental values.

\begin{figure*}
  \centering
\begin{overpic}[width=8.0cm,height=6.0cm,angle=0]{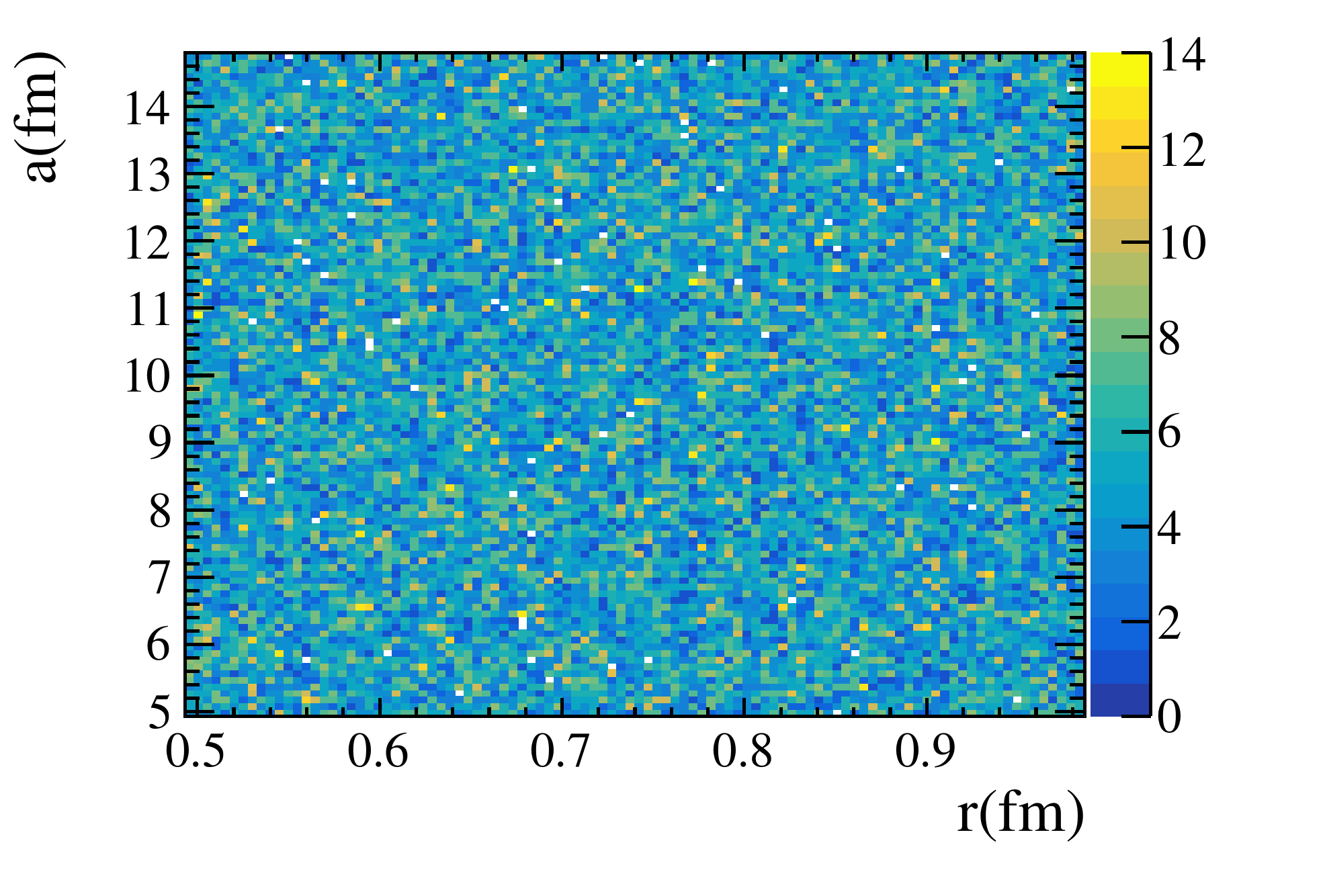}
\put(0,72){$(a)$}
\end{overpic}
 \begin{overpic}[width=8.0cm,height=6.0cm,angle=0]{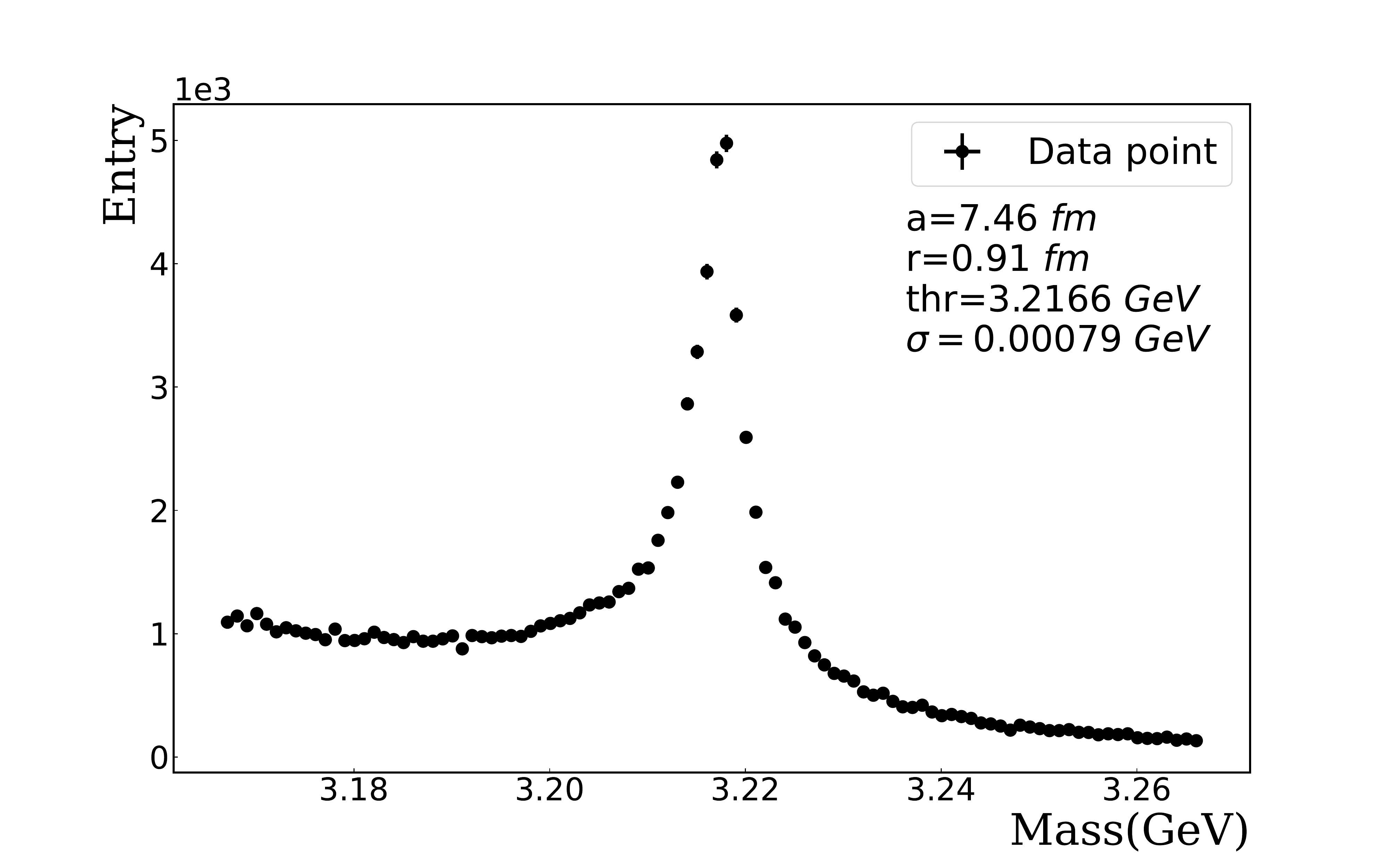}
\put(0,72){$(a)$}
\end{overpic} \\

\begin{overpic}[width=8.0cm,height=6.0cm,angle=0]{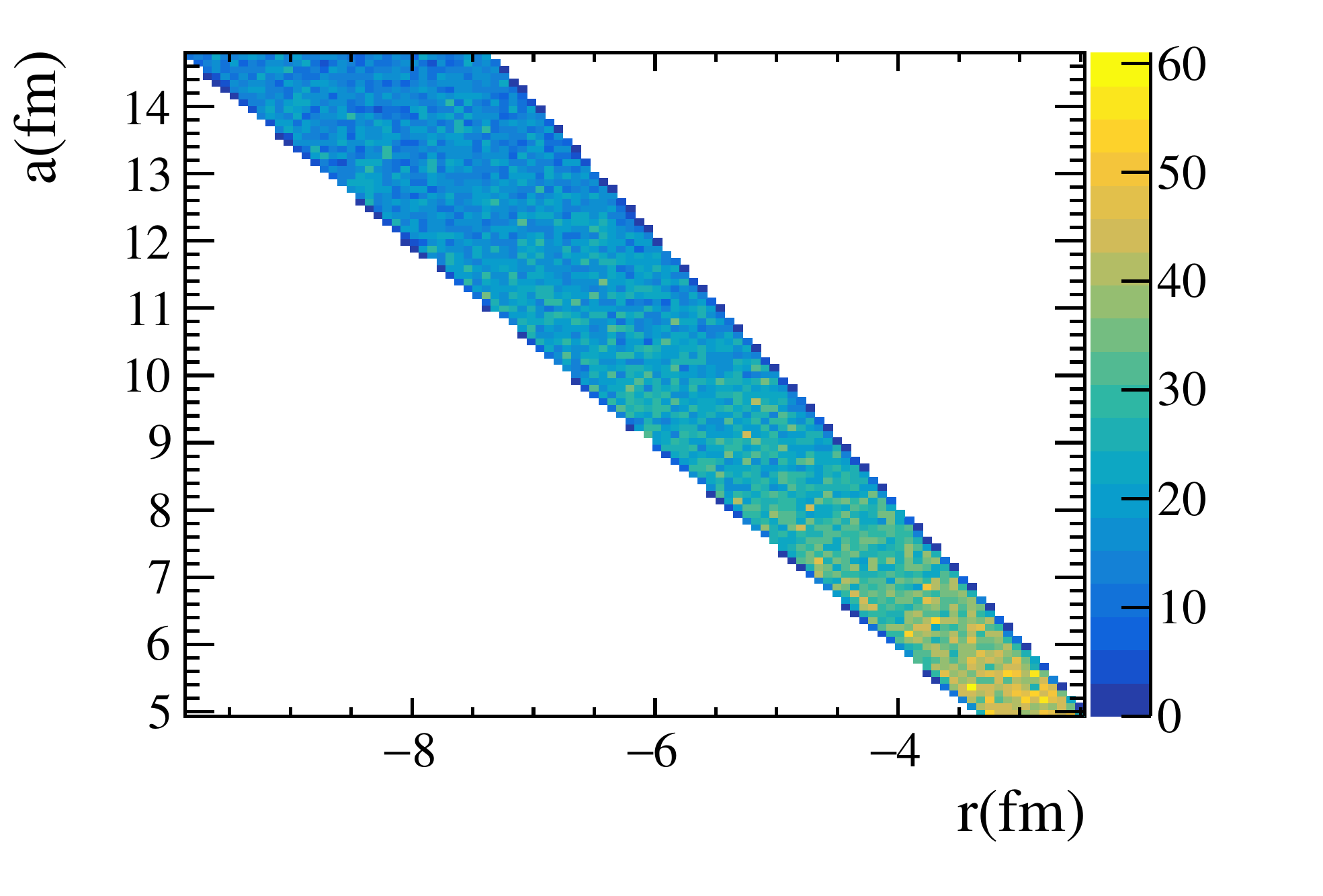}
\put(0,72){$(b)$}
 \end{overpic}
\begin{overpic}[width=8.0cm,height=6.0cm,angle=0]{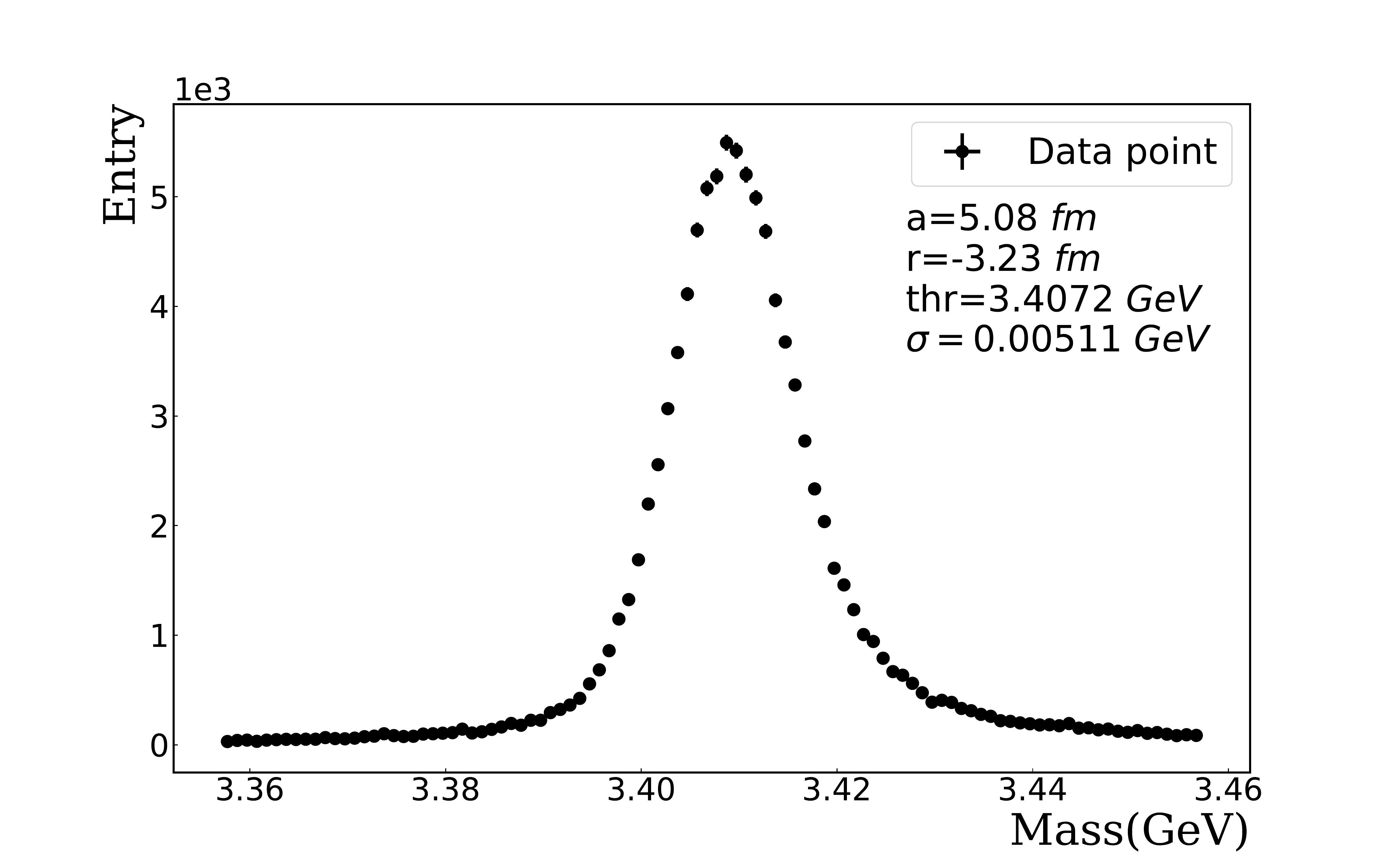}
\put(0,72){$(b)$}
 \end{overpic}\\

  \begin{overpic}[width=8.0cm,height=6.0cm,angle=0]{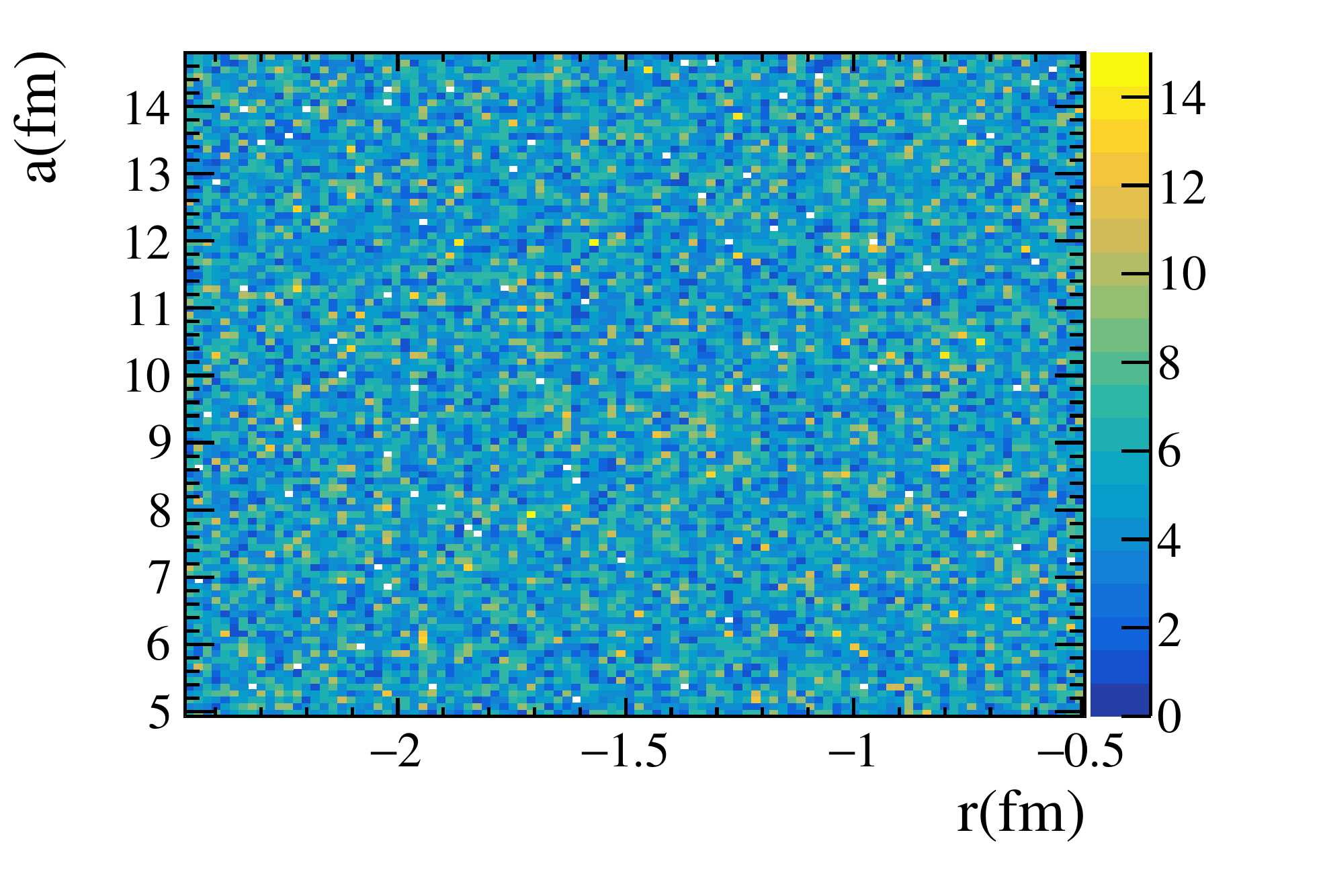}
\put(0,72){$(c)$}
 \end{overpic}
     \begin{overpic}[width=8.0cm,height=6.0cm,angle=0]{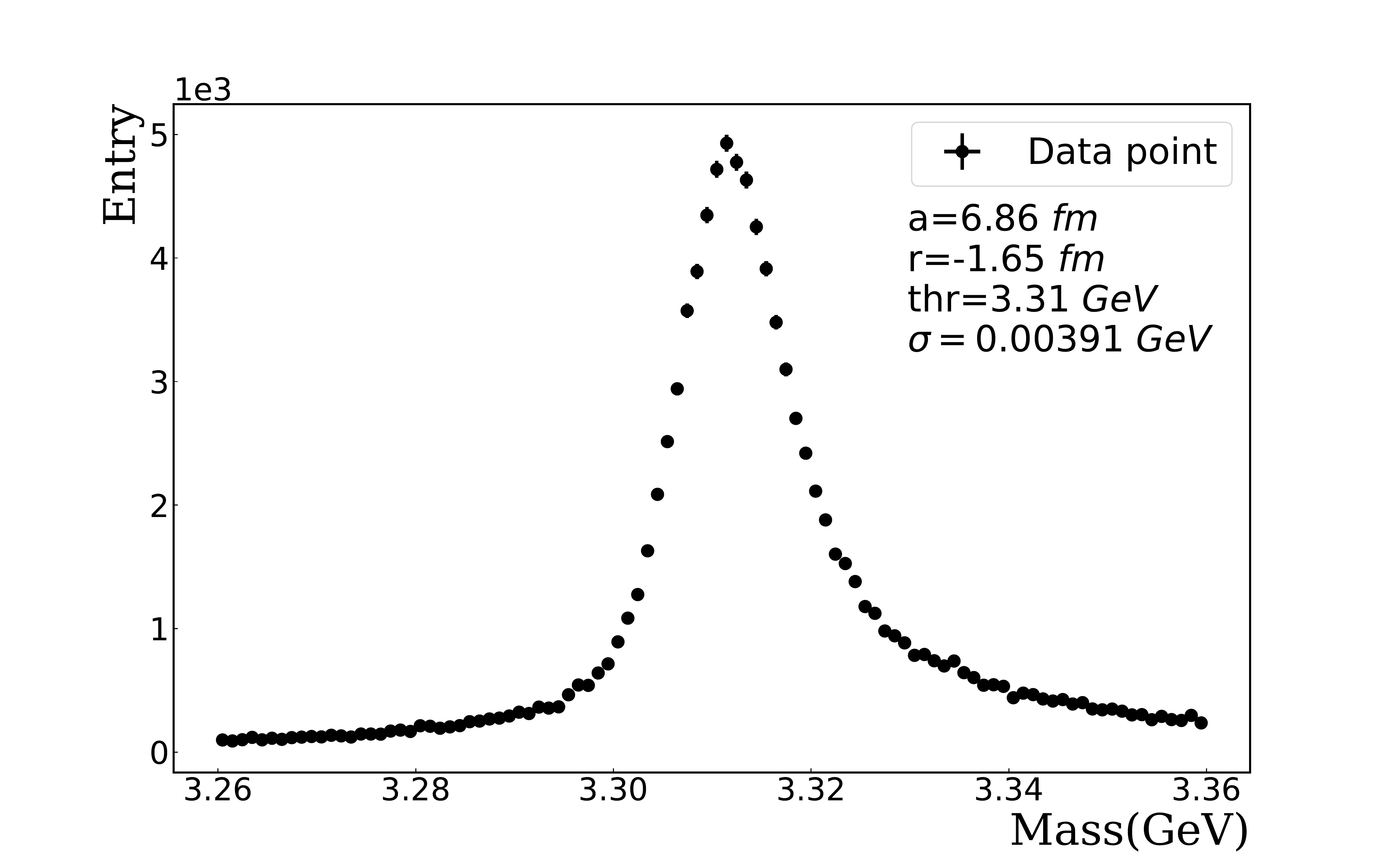}
\put(0,72){$(c)$}
 \end{overpic}

  \caption{ The left panels show 2D histograms for the parameters a and r at generation levels in cases of a bound state (a), a resonance (b), and a virtual state (c).  The right panels show 200 data points illustrated in histograms for a bound state (a), a resonance (b), and a virtual state (c), respectively, for a specific value of parameter vectors.}
  \label{Histograms}
  \end{figure*}

The training/testing datasets are generated with the Monte Carlo technique based on the open source software ROOT~\cite{root}.
The four parameters are vectorized as
\begin{align}
\mathrm{parameters}~ \mathrm{vector} = (a, r, \mathrm{threshold}, \sigma)\label{parameters list}
\end{align}
Within the ranges of Eqs.~\eqref{eq:region1},\eqref{eq:region2},\eqref{eq:region3}, and \eqref{eq:region4}, 150000 samples of the parameter vectors and corresponding histograms are uniformly generated.
Here, 45000 samples are used for testing the performance after training.
These samples are indexed as
\begin{align}
\mathrm{datasets} = \{H_i, a_i, r_i, \mathrm{threshold}_i, \sigma_i\}, i=1, ..., 150000,
\end{align}
where $H_i$ represents a histogram hosting 100 paired values, i.e., the mass spectrum.
Figure~\ref{parameters_distribution} illustrates uniform distributions of the parameter $\sigma$ and threshold.
Figure~\ref{Histograms} illustrates 2D histograms for the parameters $a$ and $r$ ( left column).
Given a specific value of parameter vector, the example mass spectra are illustrated in
the right column of Fig.~\ref{Histograms}.

\begin{figure*}
  \centering
  \includegraphics[scale=0.4]{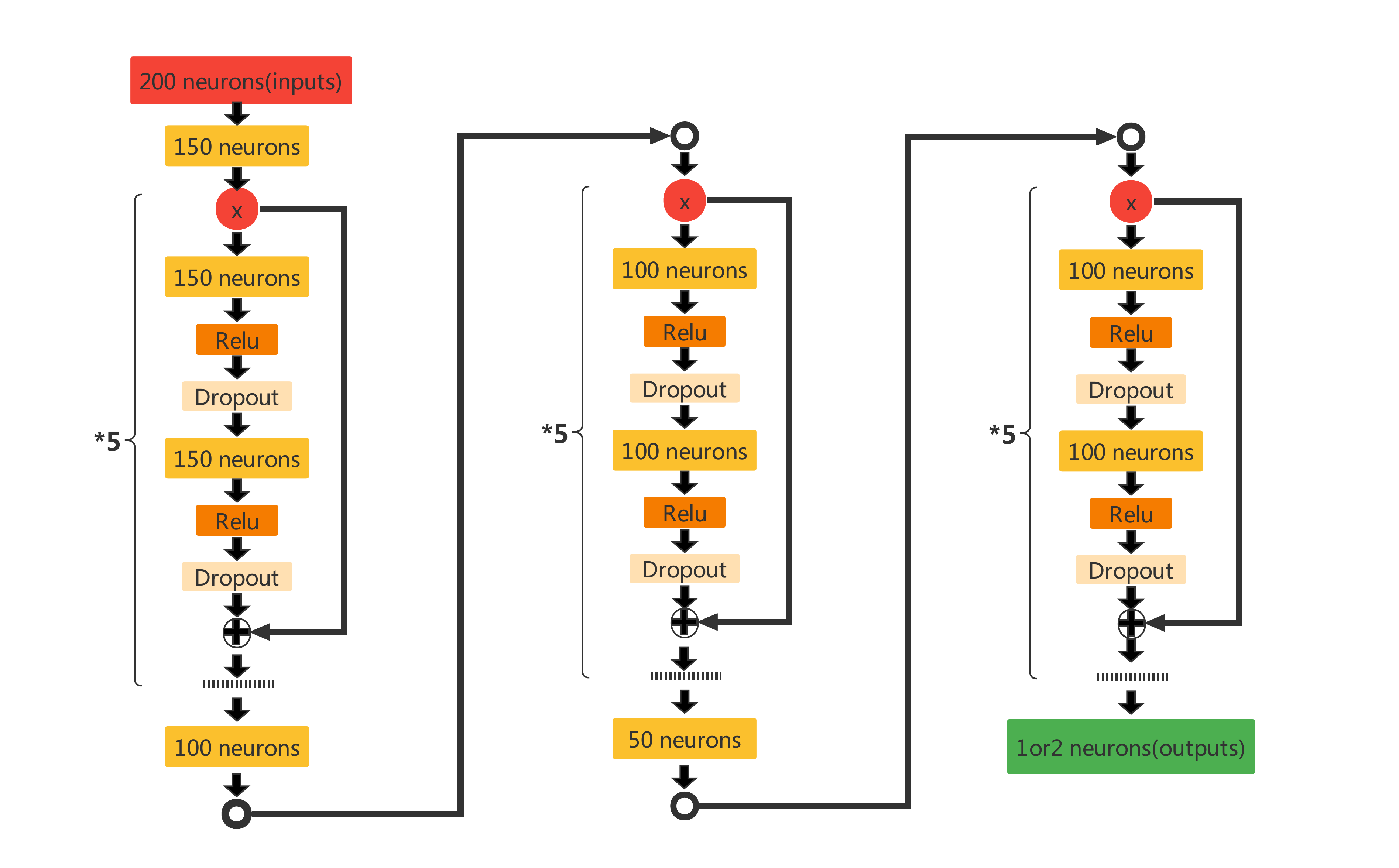}
  \caption{The structure of the ResNet-based neural network used in this work.}
  \label{ResNet}
  \end{figure*}

\section{Training}\label{training}

\begin{figure*}
  \centering
  \includegraphics[scale=0.3]{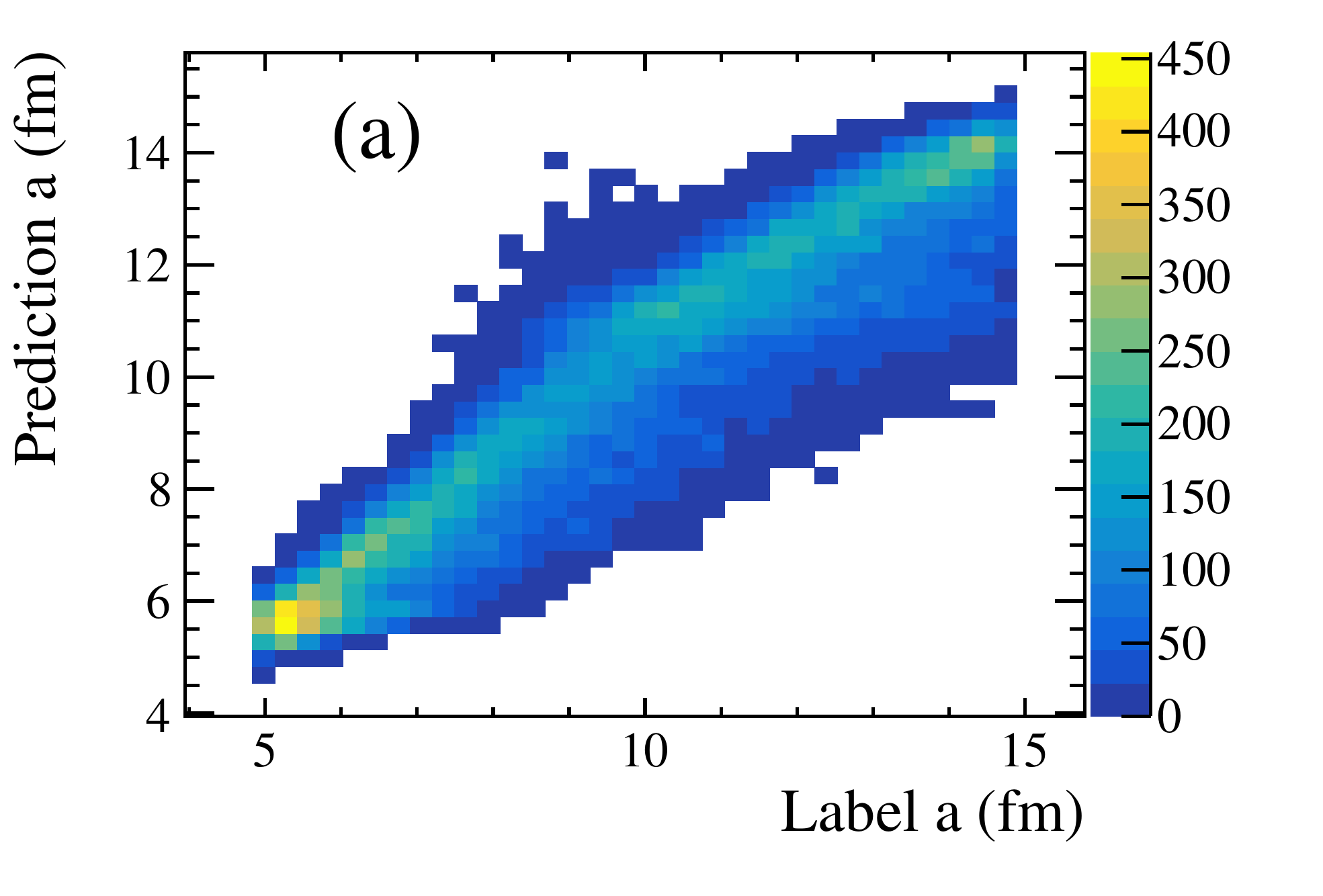}\includegraphics[scale=0.3]{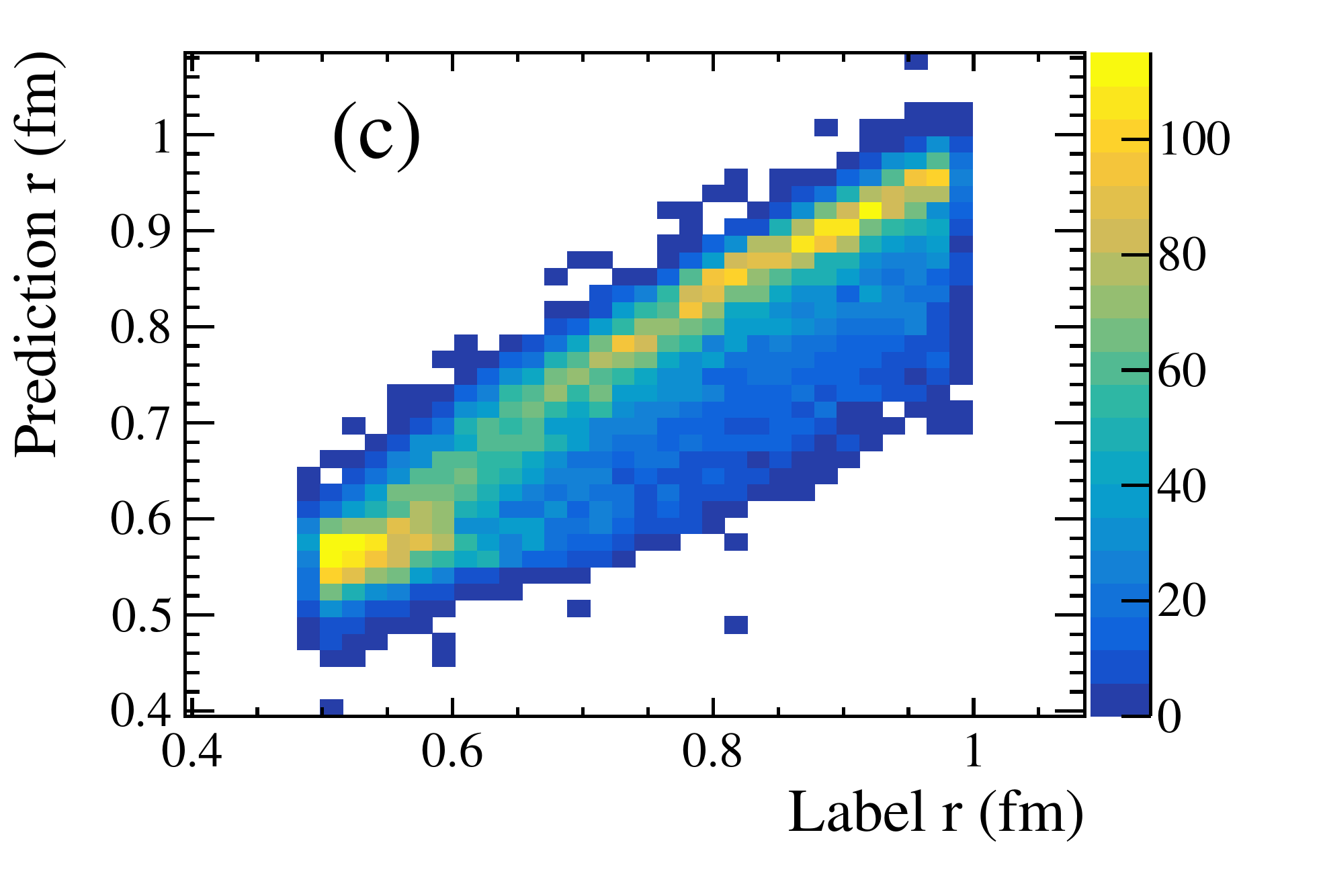}\includegraphics[scale=0.3]{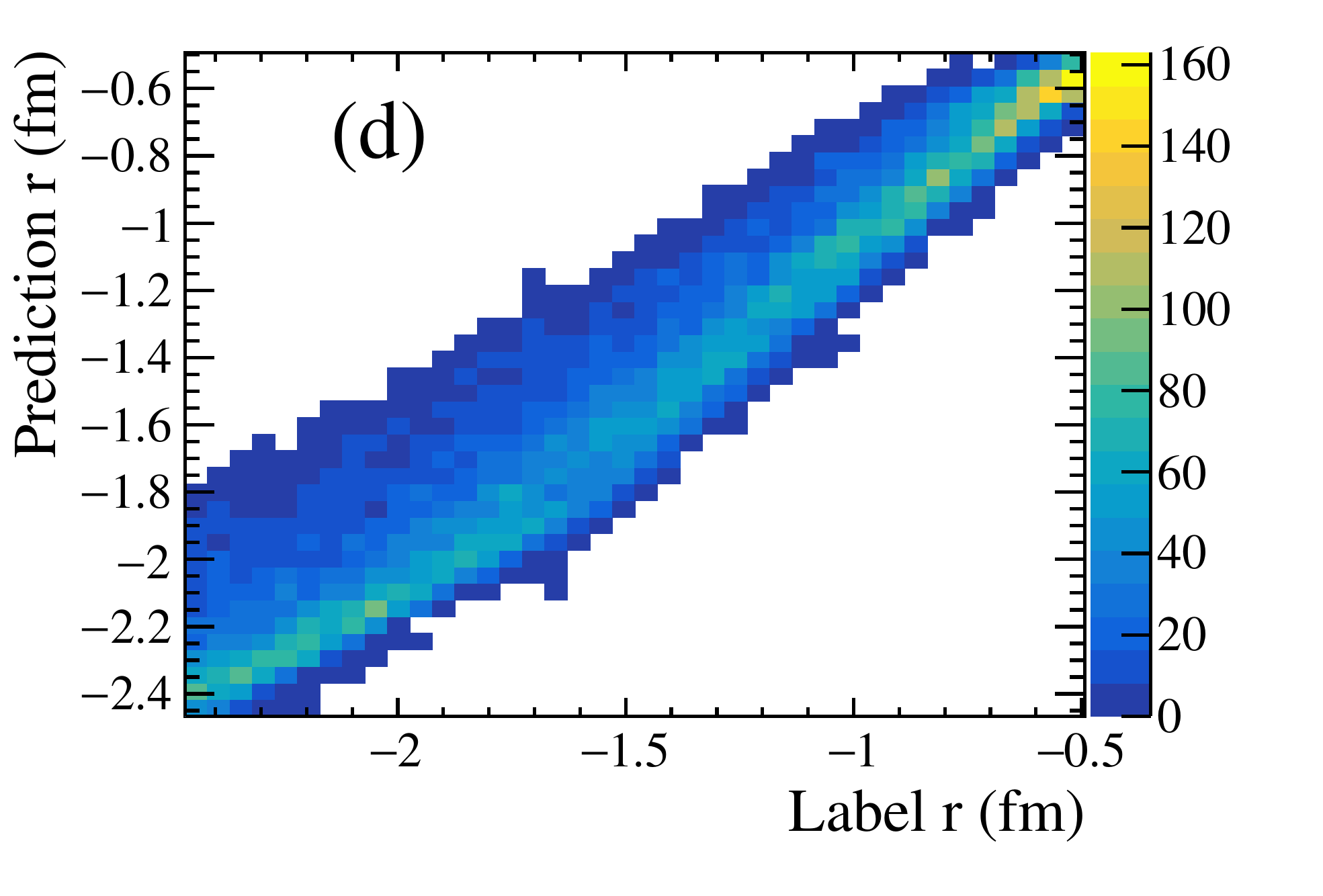}\\
  \includegraphics[scale=0.3]{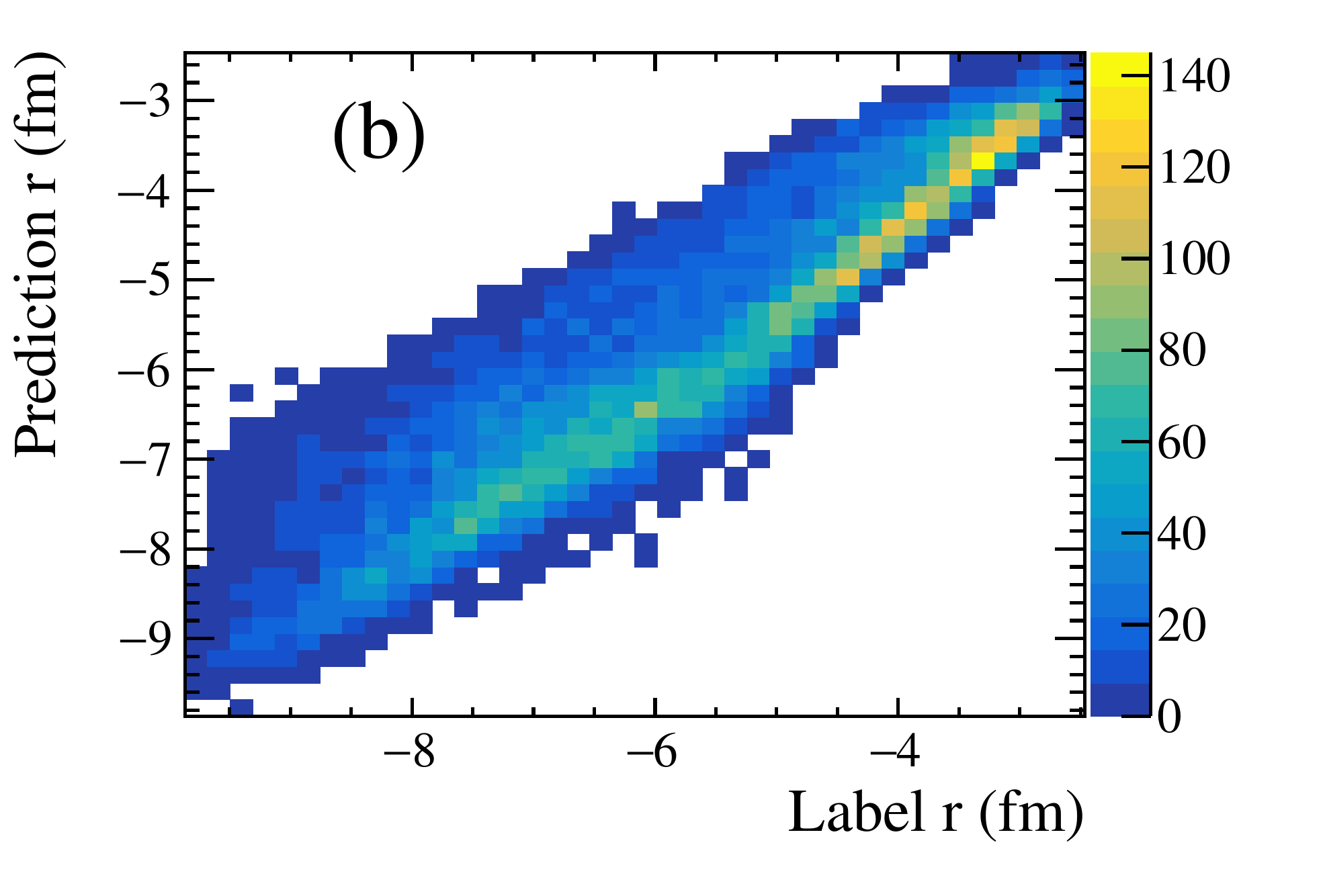}\includegraphics[scale=0.3]{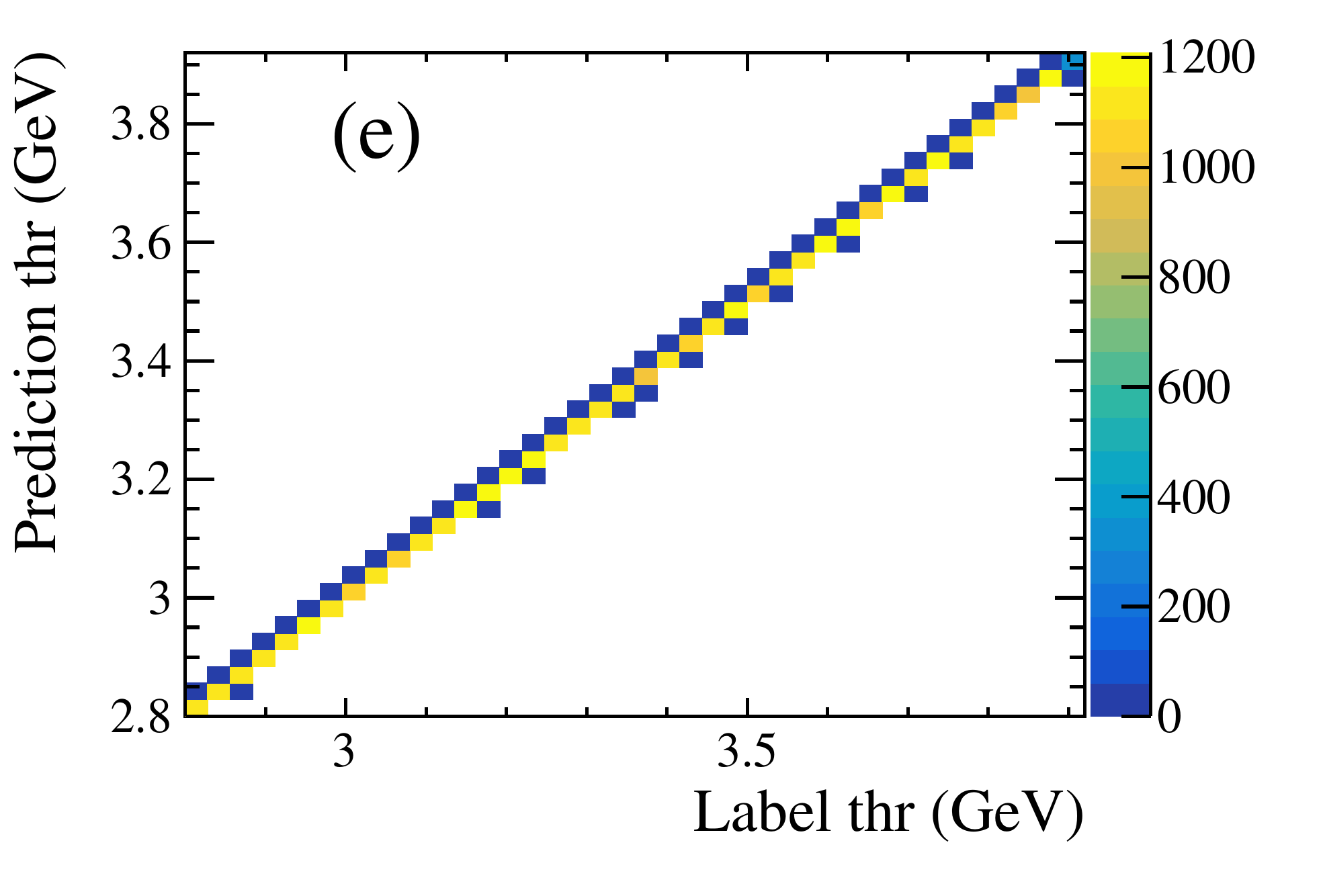}\includegraphics[scale=0.3]{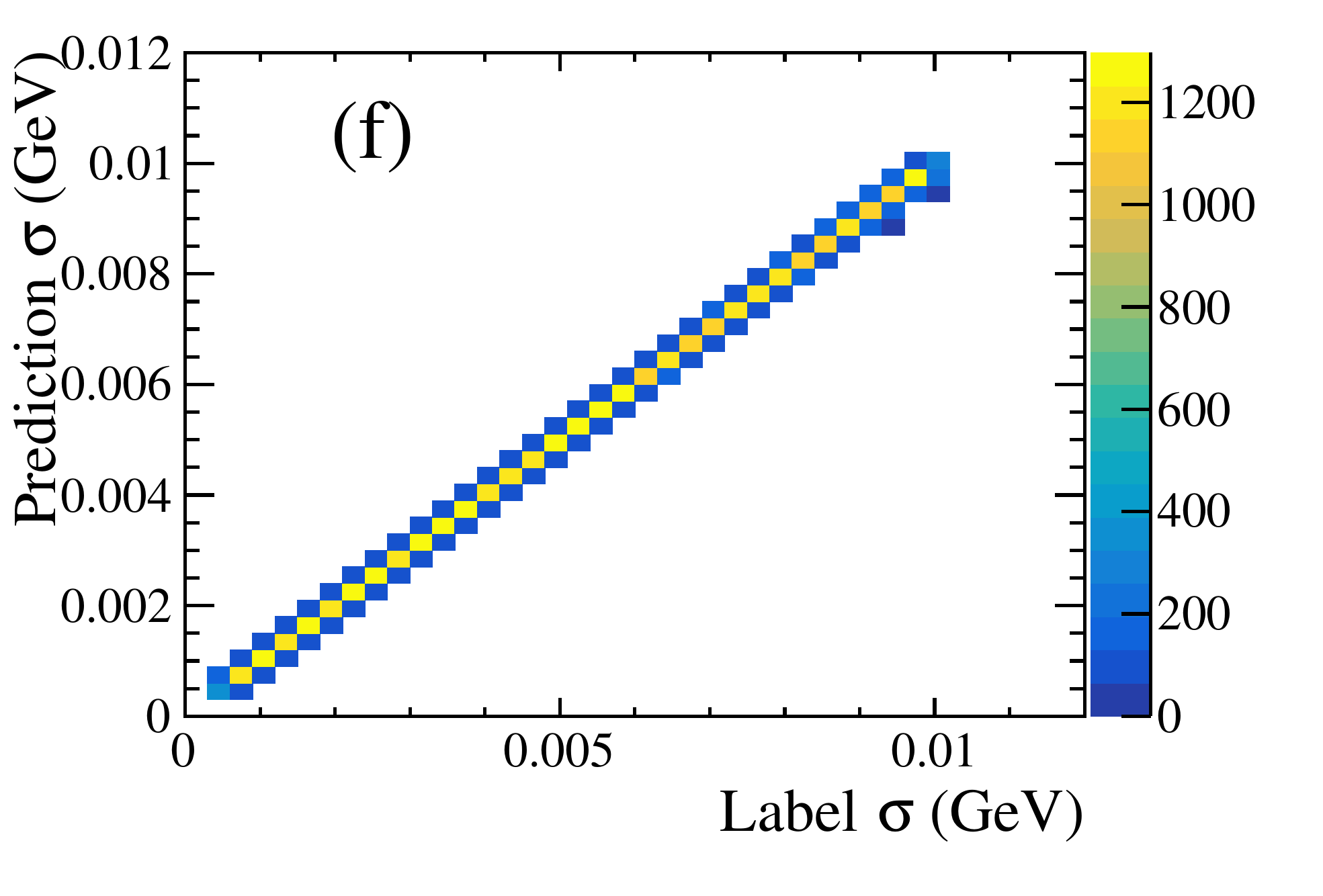}
  \caption{Correlations between predicted value and input label value.
  (a) is for scattering length $a$. (b),(c), and (d) are for effective ranges within the regions
  [0.49,0.99]~fm, [9.87,2.47]~fm, and [2.47,0.49]~fm, respectively. (e) is for
  threshold. (f) is for resolution parameter $\sigma$.}
  \label{Correlations}
  \end{figure*}

A multilayer perception~\cite{MLP}-based ResNet~\cite{ResNet} is implemented with PyTorch\cite{pytorch} to
 regress the four parameters $a$, $r$, threshold, and $\sigma$ by training the generated dataset.
The parameters $a$ and $r$ are simultaneously regressed with a model, and the other two are individually regressed  with another two models as shown in Fig.~\ref{ResNet}.
Three models are built with an identical structure,
in which the input layer is set to $200\times 1$ vector,
 followed by a dimensional reduction layer. The three ResBlocks
 compress into one or two outputs and finally connect to the parameter labels.
The ResBlock introduces a shortcut connection between the relu nonlinear activation layer and the last layer of the block .
In this way, solving the models with the Adam~\cite{adam} optimizer is of high efficiency
if we choose an optimization metric as the mean squared error function.
A reasonable solution could be obtained around 1000 training epochs using an initial learning rate value of 0.001
 and randomizing the neuron weights with a normal distribution while setting the neuron bias to zero.
Note that our labeled values have been applied with normalization and nondimensionalization.
The goodness of a solution can be measured by the correlation coefficients,
i.e., Fig.~\ref{Correlations}, which are around one indicating the equivalence of the predicted values and the labeled values.

\begin{figure}
    \centering
\includegraphics[scale=0.16]{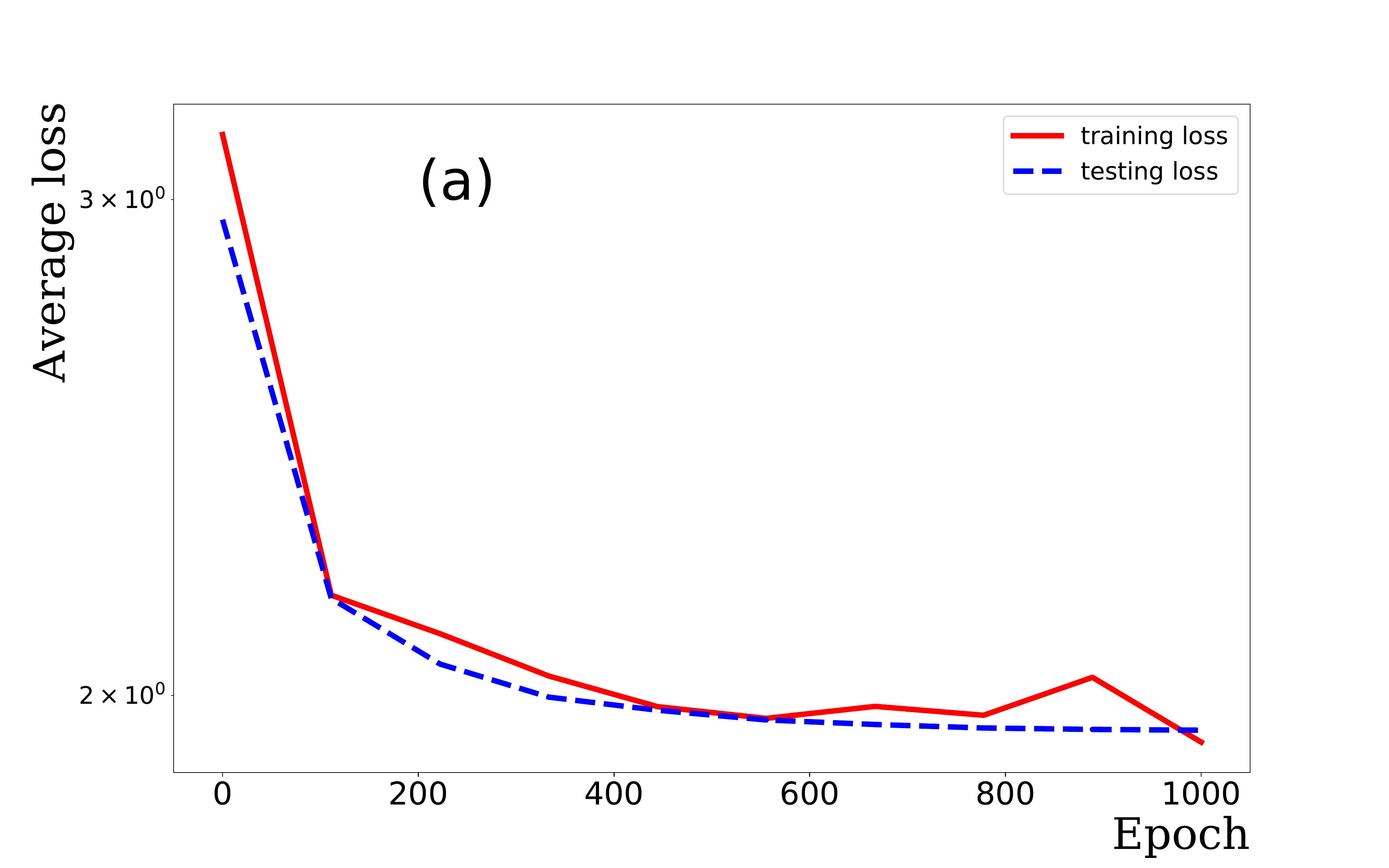}
\includegraphics[scale=0.16]{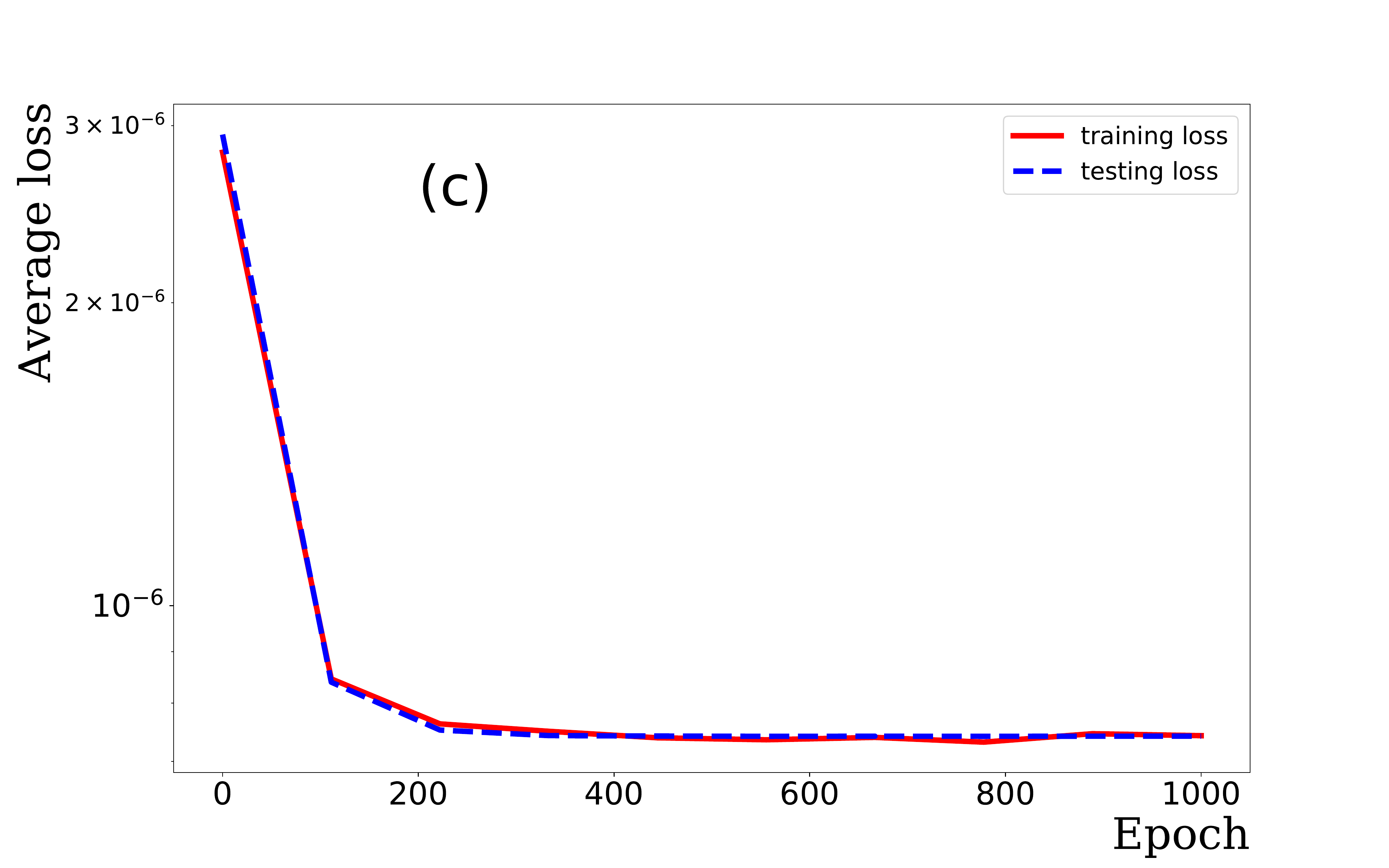}
\includegraphics[scale=0.16]{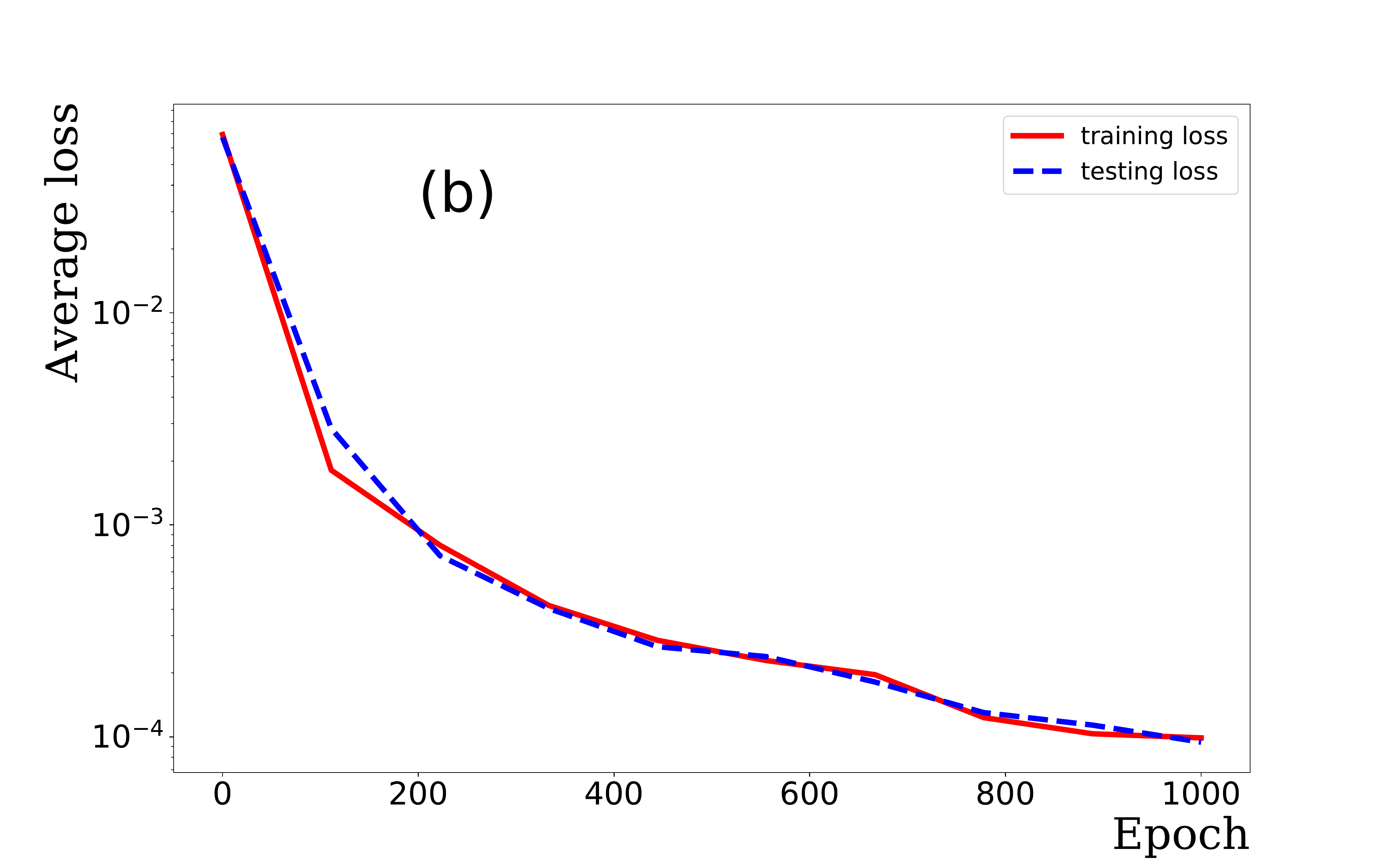}
  \caption{The MSELoss functions of scattering length/effective range (a), threshold (b), and resolution $\sigma$ (c) converge as training epoches increase. }
  \label{AbsoluteSupp}
  \end{figure}

At the beginning of training, the model needs to be initialized.
The weights of neurons are randomly initialized with a normal distribution,
 and the biases of neurons are set to zero.
The threshold values of dropout layers are set to 0.3.
The label values for the parameters $a$ and $r$ are applied with dimensionless normalization,
\begin{eqnarray}
a_{\mathrm{norm}} &=& \frac{a_{\mathrm{generation}}}{a_{\mathrm{max}}}, \\
r_{\mathrm{norm}} &=& \frac{r_{\mathrm{generation}}-r_{\mathrm{min}}}{r_{\mathrm{max}} - r_{\mathrm{min}}},
\end{eqnarray}
where $a_\mathrm{generation}$ and $r_\mathrm{generation}$ are the generated values.
$a_\mathrm{max}$, $r_\mathrm{max}$, and $r_\mathrm{min}$ are their maximum and minimum values.
These two parameters are simultaneously regressed because they are largely correlated for given cases, while the threshold and $\sigma$ are individually regressed since they are independent.
To solve our model, the Adam~\cite{adam} optimizer, one of the most widely used optimizers which combines the momentum algorithm  and the RMSProp algorithm\cite{rmsprop},
 is used. It does not only fasten the convergence but also reduce the fluctuation of the loss function,
 which is defined as the MSELoss function (the mean squared error loss) to  measure the Euclidean distance between the prediction values and the label values.
A reasonable solution can be achieved by using around 1000 training epochs with
 an initial learning rate value of 0.001,  which is automatically and dynamically adjusted during the training cycle.
As illustrated in Fig.~\ref{AbsoluteSupp}, the MSELoss function converges rapidly after 200 iterations for the regression.

 \begin{figure*}
\centering
\includegraphics[scale=0.45]{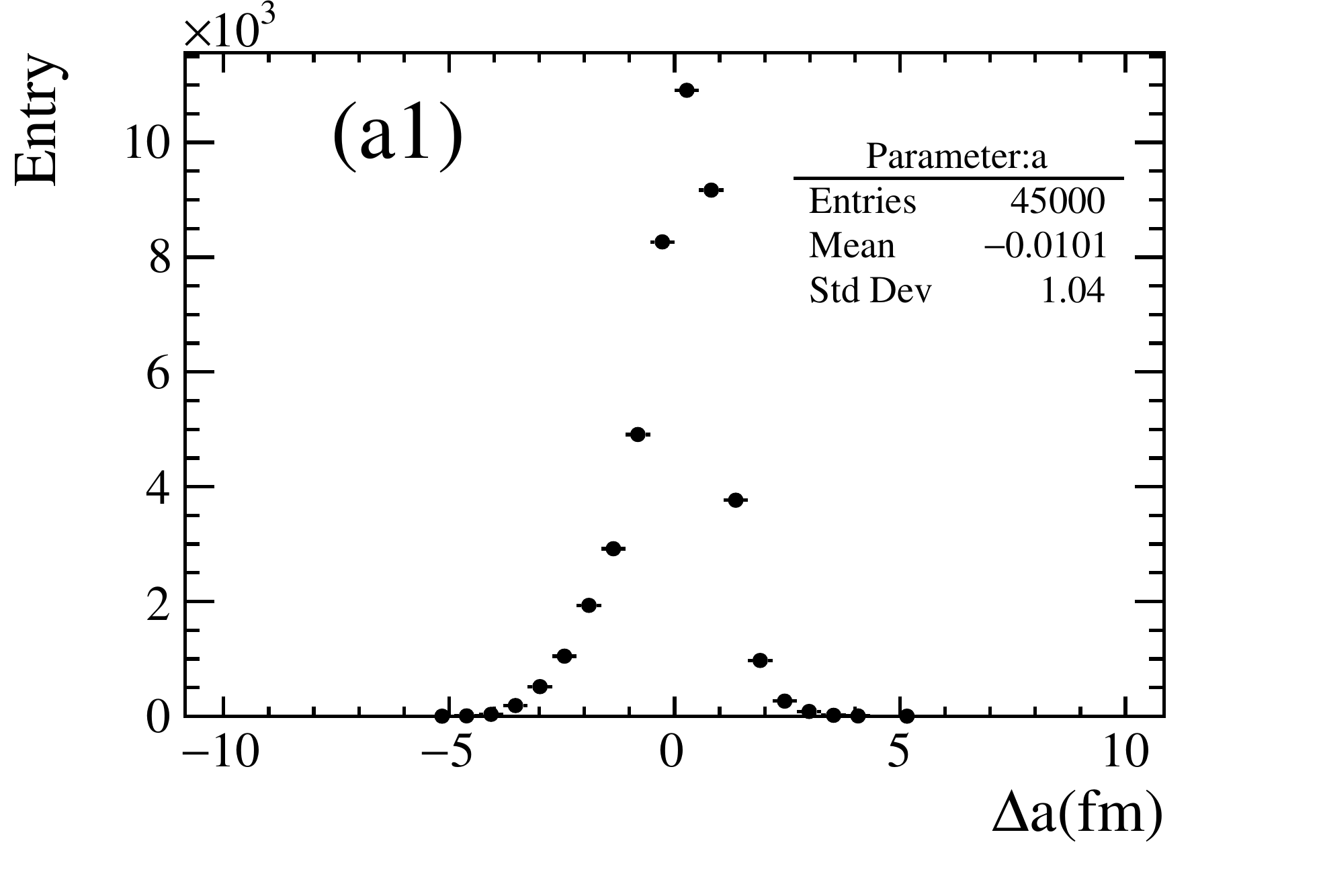}~~~\includegraphics[scale=0.45]{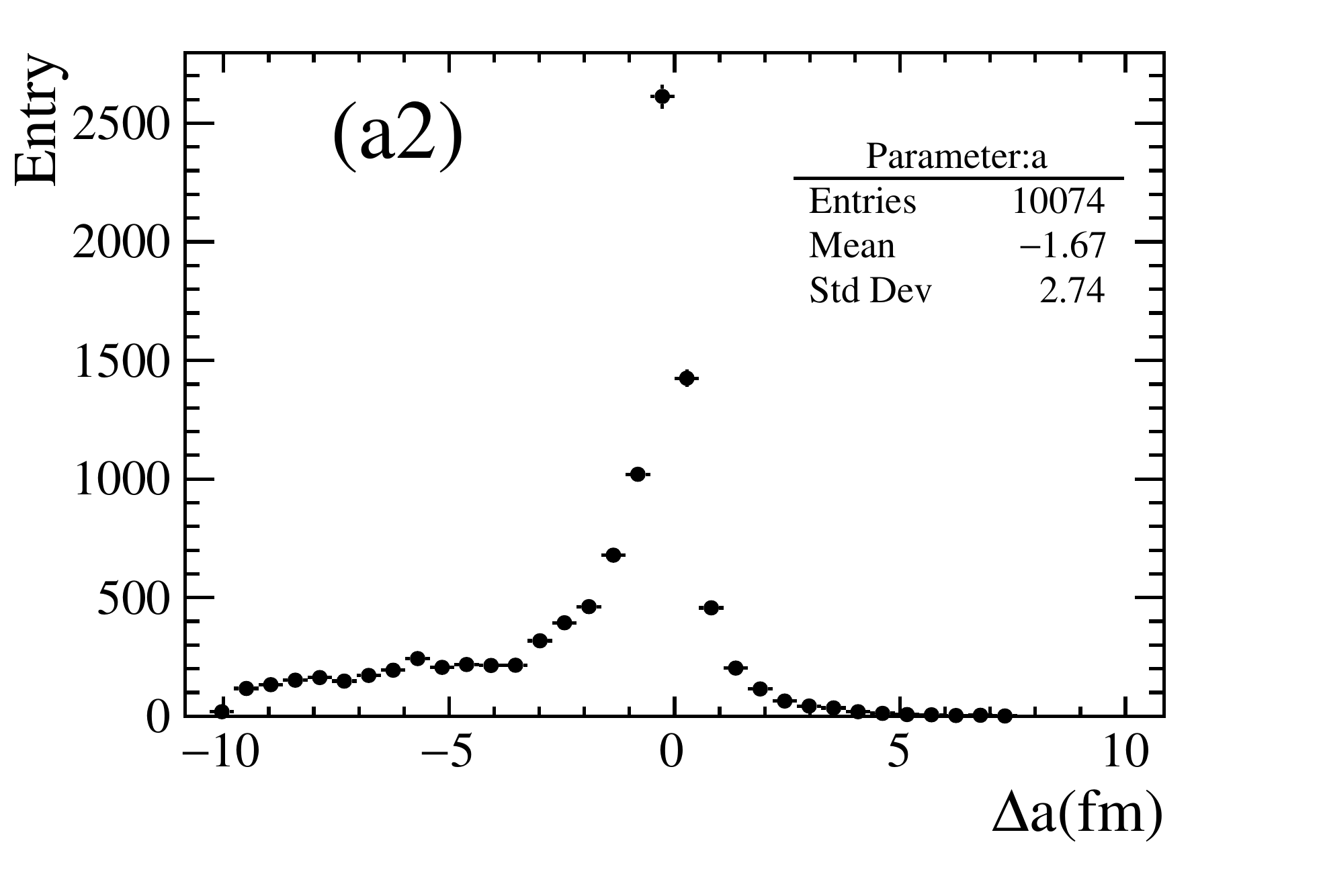} \\
\includegraphics[scale=0.45]{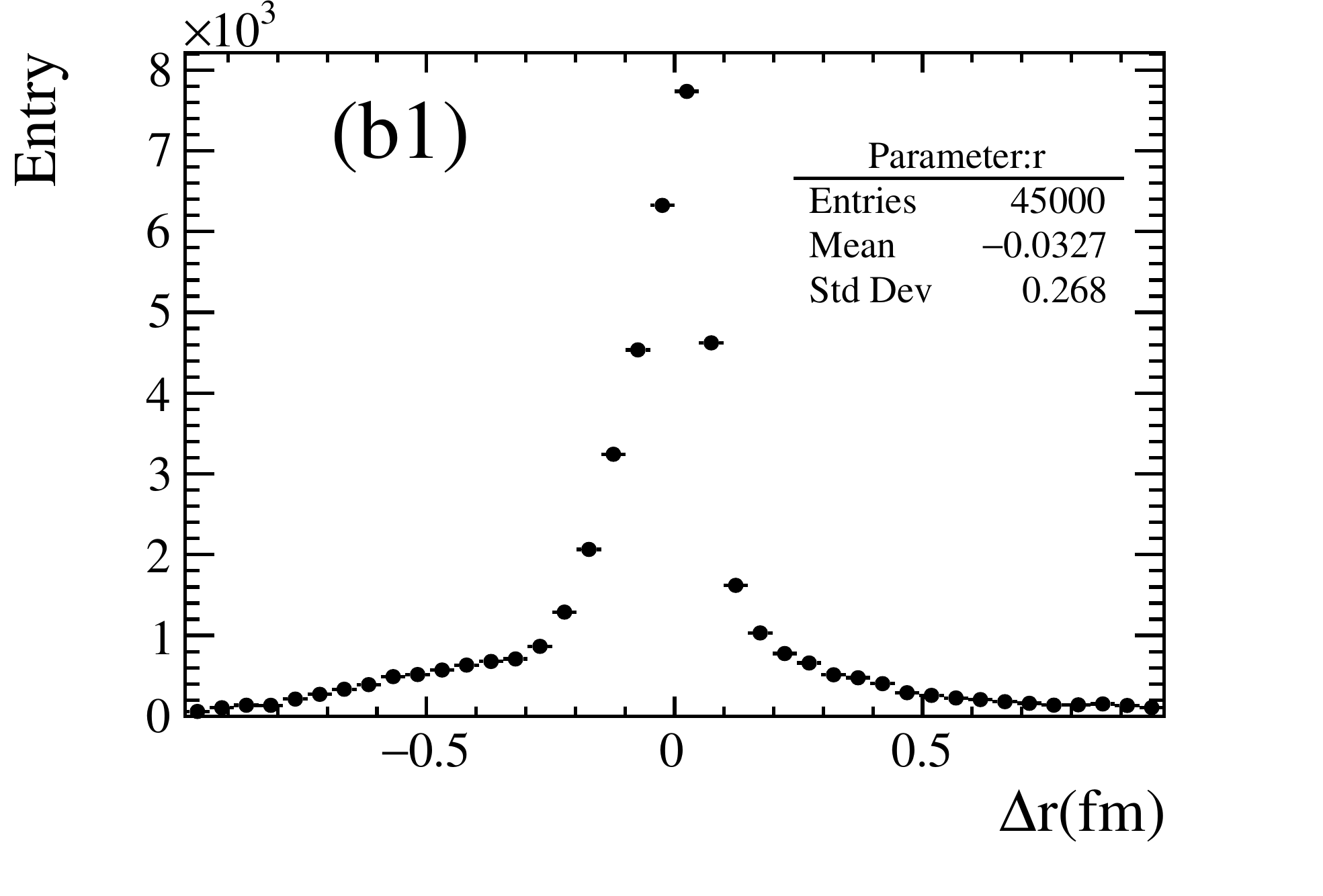}~~~\includegraphics[scale=0.45]{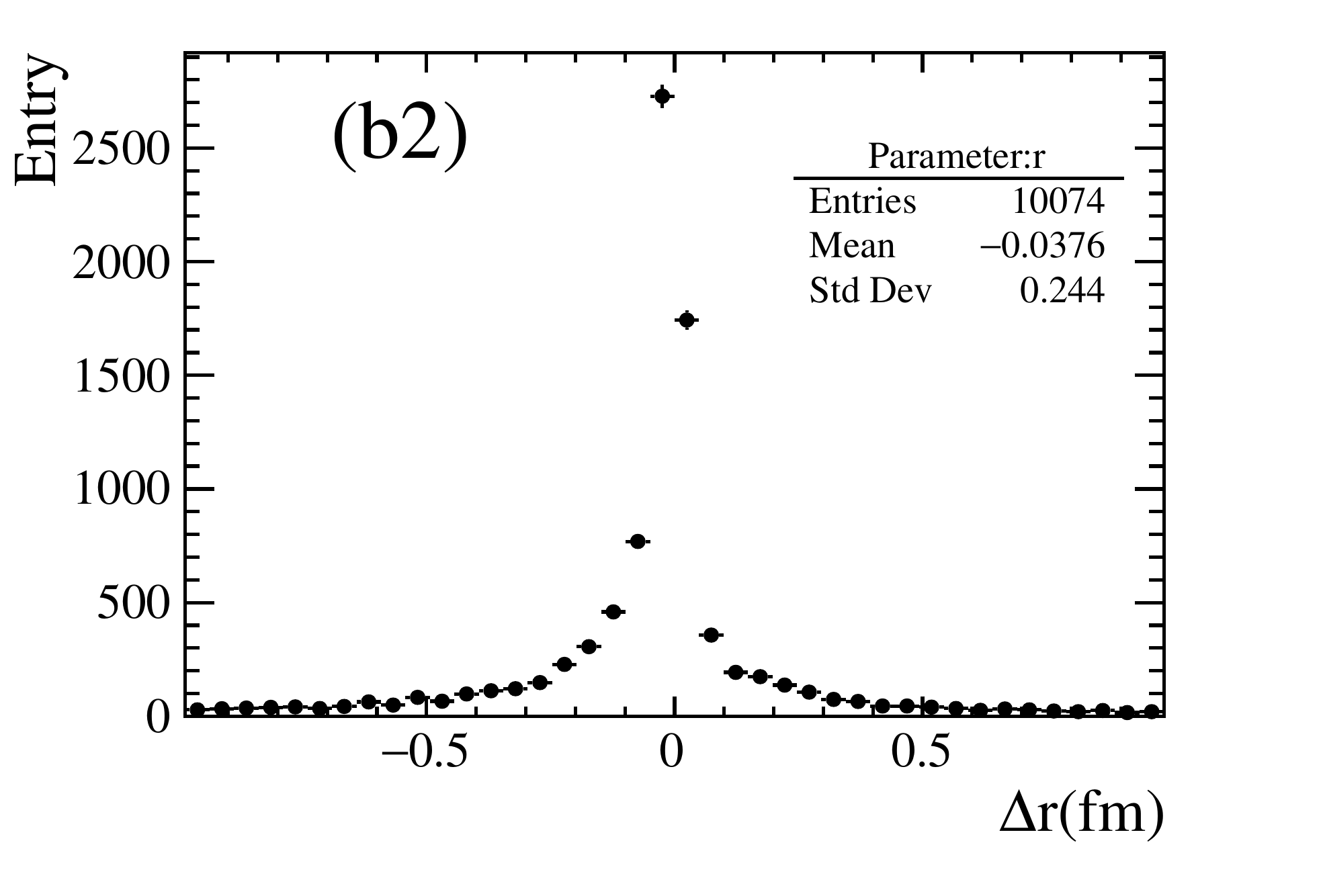} \\
\includegraphics[scale=0.45]{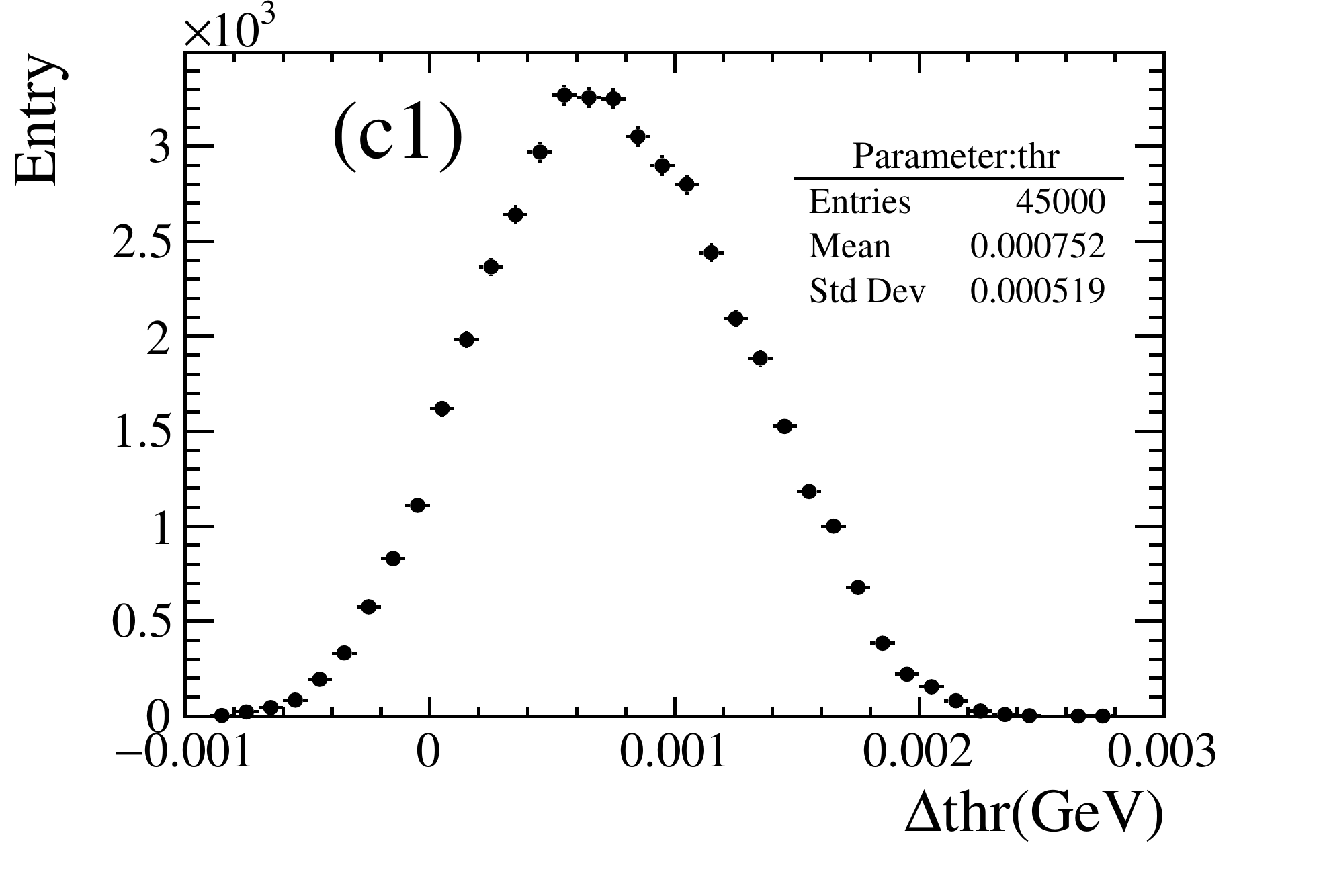}~~~\includegraphics[scale=0.45]{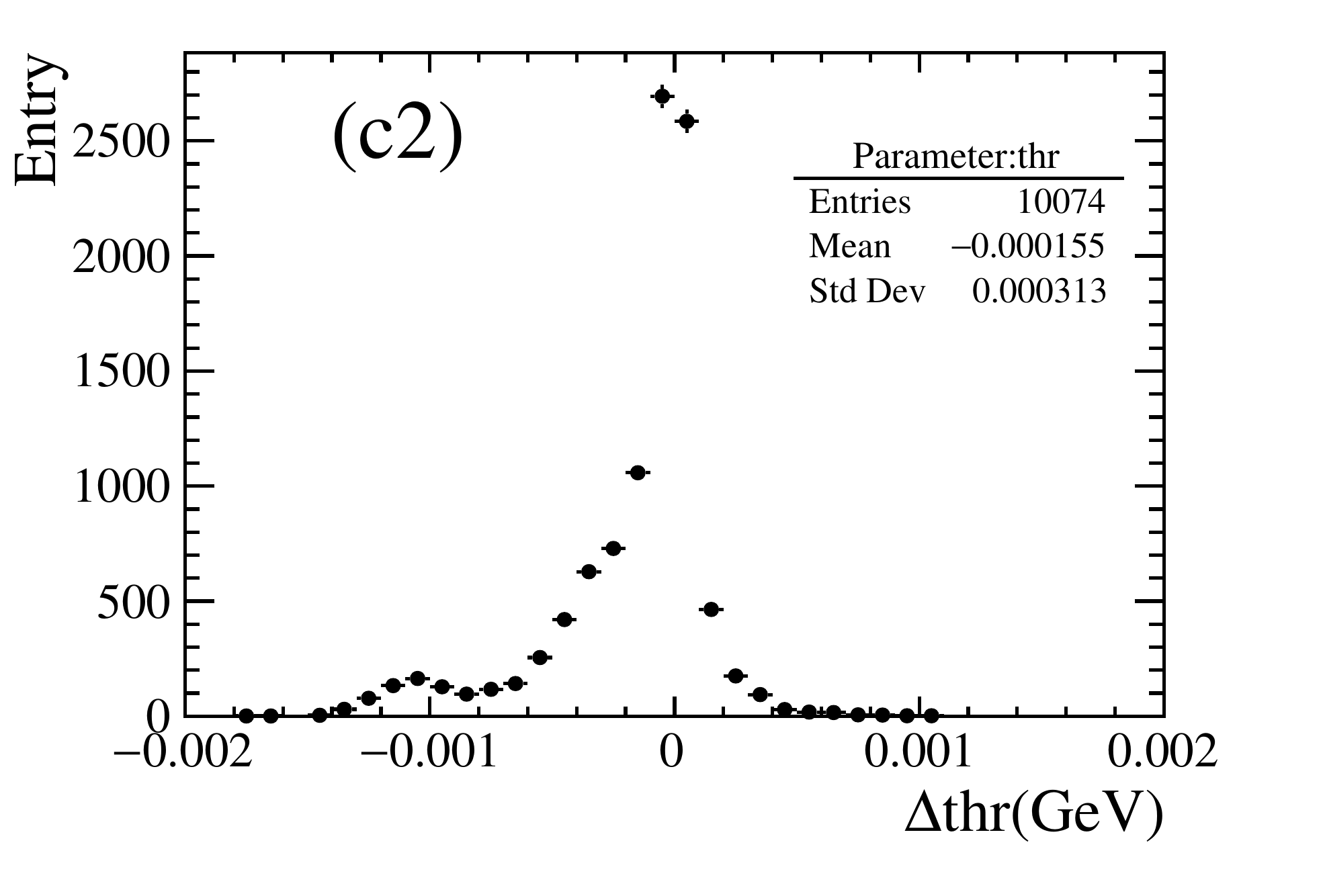} \\
\includegraphics[scale=0.45]{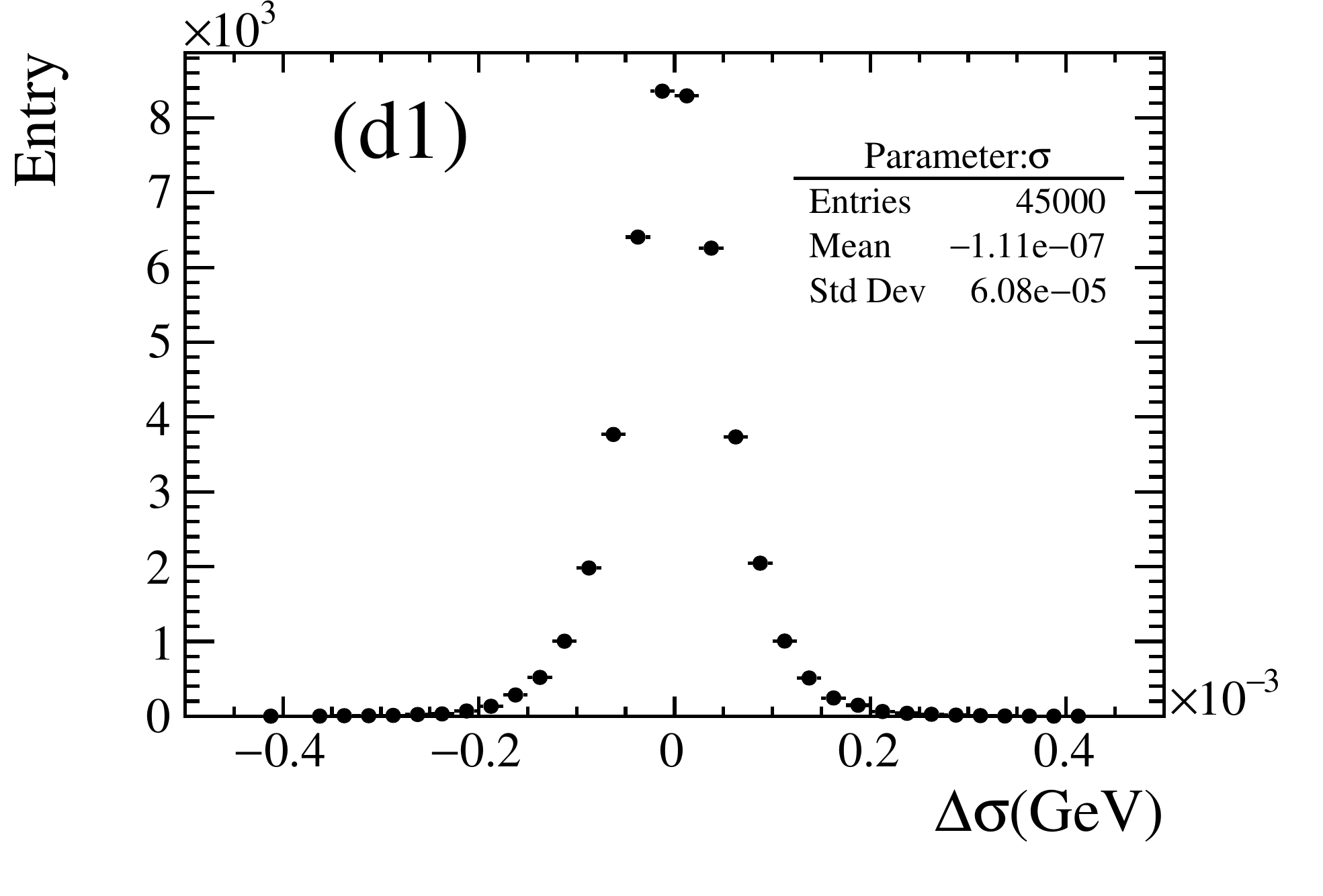}~~~\includegraphics[scale=0.45]{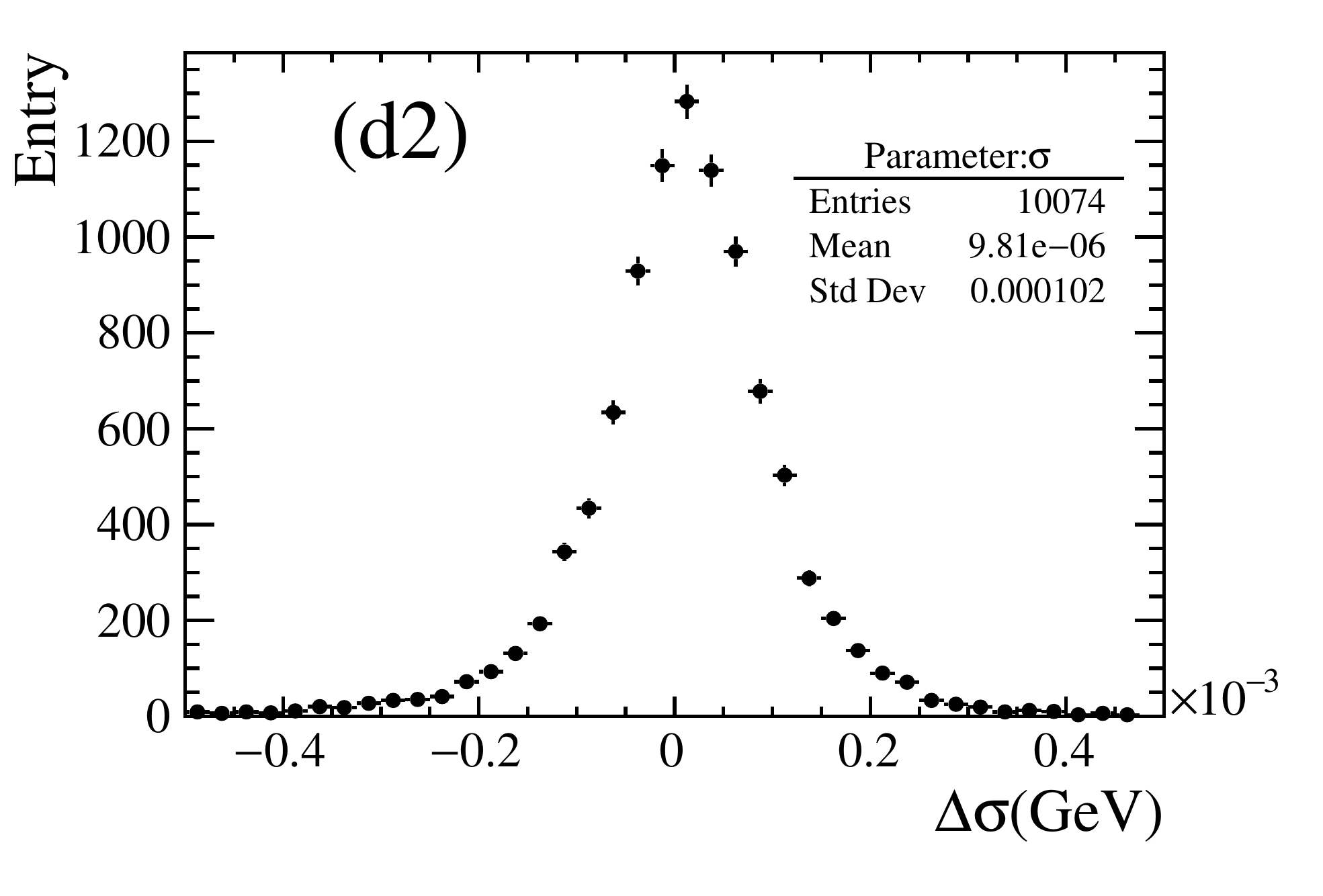}
\caption{Distributions of the difference between prediction values and label values for each parameter. Distributions in the left column are obtained with deep learning while those in the right column with the fitting method.}
\label{fig:biaserrors}
\end{figure*}

\section{Evaluation}
We further extract the distribution of the difference between the predicated values and the label values as shown in Fig.~\ref{fig:biaserrors} (left column), in which plots (a1–d1) are for the parameters $a, r$, threshold, and $\sigma$.
These distributions are obtained by testing 45000 samples.
The mean measures the deviation of the predicted values from the labeled ones.
The root-of-mean-square (RMS) measures the intrinsic uncertainties of this method.
Relevant numbers are summarized in the Table~\ref{biaserrorx}.
For a straightforward comparison, we extract the parameters from directly fitting to our testing samples.
Among the 15000 samples, only 10074 fitting give acceptable $\chi^2$s, i.e., less than 100.
For those successful fittings, the distributions of the differences between the fitted parameters
and their real values are shown in the right column of Fig.~\ref{fig:biaserrors}.
We have found that biases of the deep learning could be neglected for all parameters,
and intrinsic uncertainties could be neglected for the parameters $a$, threshold and $\sigma$.

\begin{table}[htpb]
\begin{center}
\renewcommand{\tablename}{Table}
\caption{The biases and errors information of models}\label{biaserrorx}
	\begin{tabular}{|l|c|c|c|c|}
	\hline
	           Methods$\to$    &   \multicolumn{2}{c|}{Deep learning}  & \multicolumn{2}{c|}{  Fitting}   \\
\hline
       Parameters$\downarrow$            & Bias       & Uncertainty & Bias & Uncertainty \\
                   \hline
      $a$ (fm)         &  0.010 &  1.040 & 1.67  &  2.740\\
     $r$ (fm)         &  0.033 & 0.268 &  0.038 &  0.244 \\
      threshold (MeV)        & 0.75  & 0.52 &  0.16  & 0.31 \\
      $\sigma$ (MeV)     & 0.0001 & 0.06 &  0.0098 & 0.10\\
	\hline
	\end{tabular}
\end{center}
\end{table}

\section{Apply to the $X(3872)$ and the $T_{cc}^+$}
During the last decades, tens of exotic candidates have been reported
~\cite{Chen:2016qju,Chen:2016spr,Dong:2017gaw,Lebed:2016hpi,Guo:2017jvc,Liu:2019zoy,Albuquerque:2018jkn,Yamaguchi:2019vea,Guo:2019twa,Brambilla:2019esw}.
Among them, the first and most interesting one is the $X(3872)$
which was reported by Belle Collaboration in 2003~\cite{Belle:2003nnu}. Intensive studies
have been put forward to understand its nature. For instance,
the popular explanations are the $D\bar{D}^*+c.c.$ hadronic molecule,
compact tetra-quark, and the normal charmonium with the mixture of the $D\bar{D}^*+c.c.$ hadronic molecule.
For the detailed discussions, we refer to Refs.
~\cite{Chen:2016qju,Chen:2016spr,Dong:2017gaw,Lebed:2016hpi,Guo:2017jvc,Liu:2019zoy,Albuquerque:2018jkn,Yamaguchi:2019vea,Guo:2019twa,Brambilla:2019esw}.
The first two scenarios can be distinguished by the
pole counting near the $D\bar{D}^*+c.c.$ threshold,
i.e., two poles and one pole for compact
and hadronic molecules~\cite{Guo:2017jvc}, respectively. These pole positions
are largely related to the values of scattering length and effective range.
Thus, extracting these two values could help to shed light on the nature of exotic hadrons.
Besides the $X(3872)$, another interesting one is $T_{cc}^+$~\cite{LHCb:2021vvq,LHCb:2021auc} reported by LHCb in the $D^0D^0\pi^+$ channel.
Since it is very close to the $D^{*+}D^0$ and $D^+D^{*0}$ channels,
it is viewed as a partner of the $X(3872)$ in the molecular picture.
 In the isospin limit, i.e. neglecting the mass differences between charged and neutral charmed mesons,
the $X(3872)$ and the $T_{cc}^+$ are only one-channel cases, i.e.,
the $D\bar{D}^*+c.c.$
and $DD^*$ channel, respectively.
Thus, we take them as an illustration of the applicability of our network.
Although, their isospin breaking effect
has several impacts on physical observables
~\cite{Meng:2021kmi,Wu:2021udi,Zhou:2017txt,Takeuchi:2014rsa,Li:2012cs,Karliner:2010sz,Gamermann:2009fv,Terasaki:2009in,Voloshin:2007hh,Tornqvist:2004qy}, as the first step,
we start from the one-channel case and check the applicability.
\begin{table}
  \begin{center}
  \renewcommand{\tablename}{Table}
  \caption{Parameters of the $X(3872)$ from deep learning (second column)
   and fit (third column) to the data directly.}
    \begin{tabular}{|c|c|c|}
    \hline
    \bm{$X(3872)$}\;\textbf{parameters}  & Deep learning & Fit\\
    \hline
    Parameter $a$ (fm) & $8.76\pm 1.75$ & $9.95\pm 0.34$ \\
    \hline
    Parameter $r$ (fm) & $0.56\pm 0.55$ & $0.32\pm  0.08$ \\
    \hline
    Parameter threshold (MeV) & $3871.30\pm 0.52$ & $3871.20\pm 0.01$ \\
    \hline
    Parameter $\sigma$ (MeV) & $1.20\pm 0.15$ & $1.70\pm 0.16$\\
    \hline
    \end{tabular}
  \label{X3872 prediction}
  \end{center}
  \end{table}
\begin{table}
\begin{center}
\renewcommand{\tablename}{Table}
\caption{Parameters of the $T_{cc}^+$ from deep learning (second column)
and fit (third column) to the data directly.}
	\begin{tabular}{|c|c|c|}
	\hline
	\bm{$T_{cc}^+$}\;\textbf{parameters}  & Deep learning & Fit\\
	\hline
	Parameter $a$ (fm) & $8.23\pm 1.04$ & $13.74\pm 4.77$ \\
	\hline
	Parameter $r$ (fm) & $-2.79\pm 0.27$ & $-2.15\pm 0.21$ \\
	\hline
	Parameter threshold (MeV) & $3874.83\pm 0.51$ & $3874.53\pm 0.13$ \\
	\hline
	Parameter $\sigma$ (MeV) & $1.10\pm 0.06$ & $0.11\pm 0.12$\\
	\hline
	\end{tabular}
\label{Tcc prediction}
\end{center}
\end{table}

          \begin{figure}
    \centering
\includegraphics[scale=0.2]{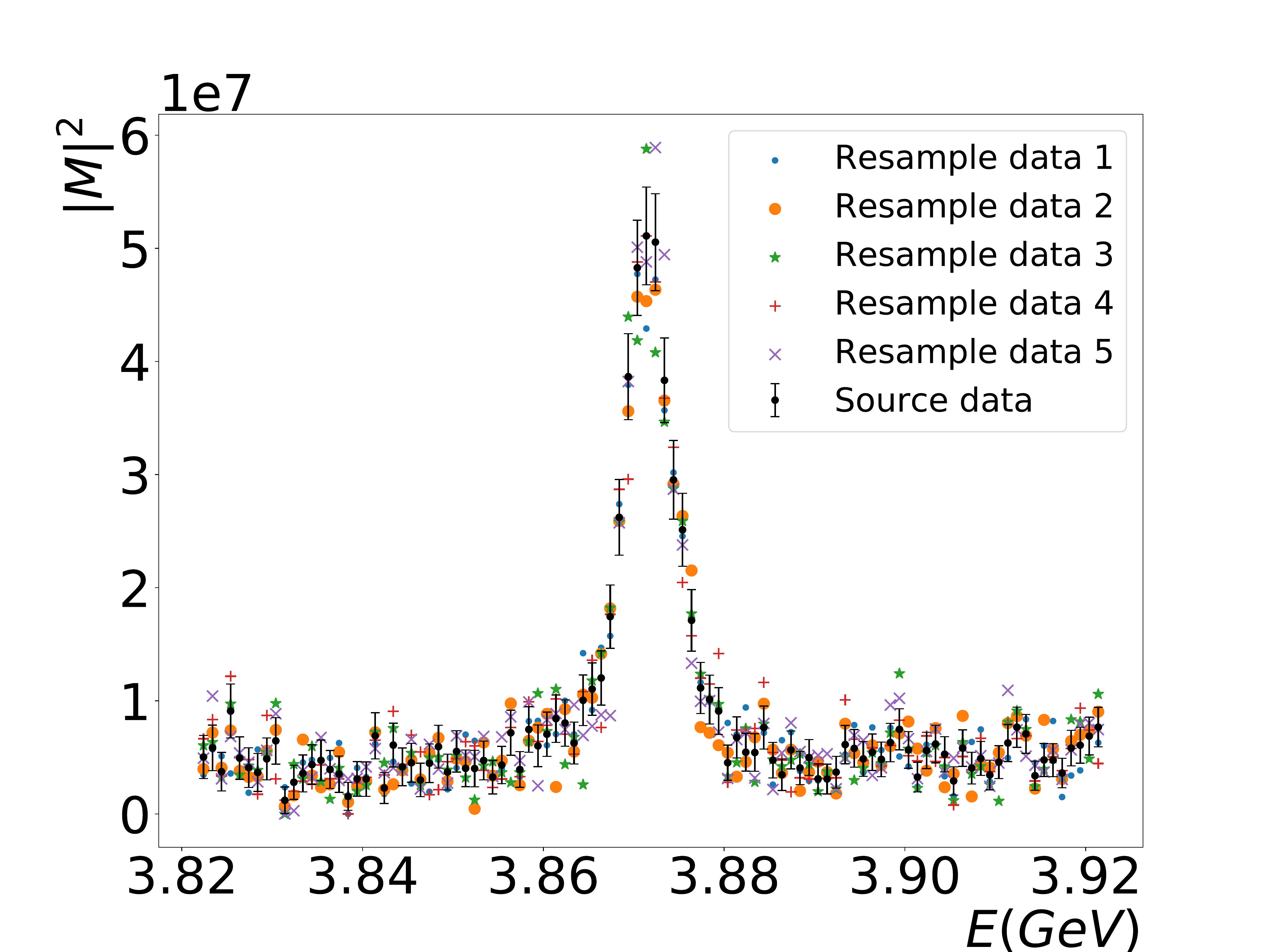}\includegraphics[scale=0.2]{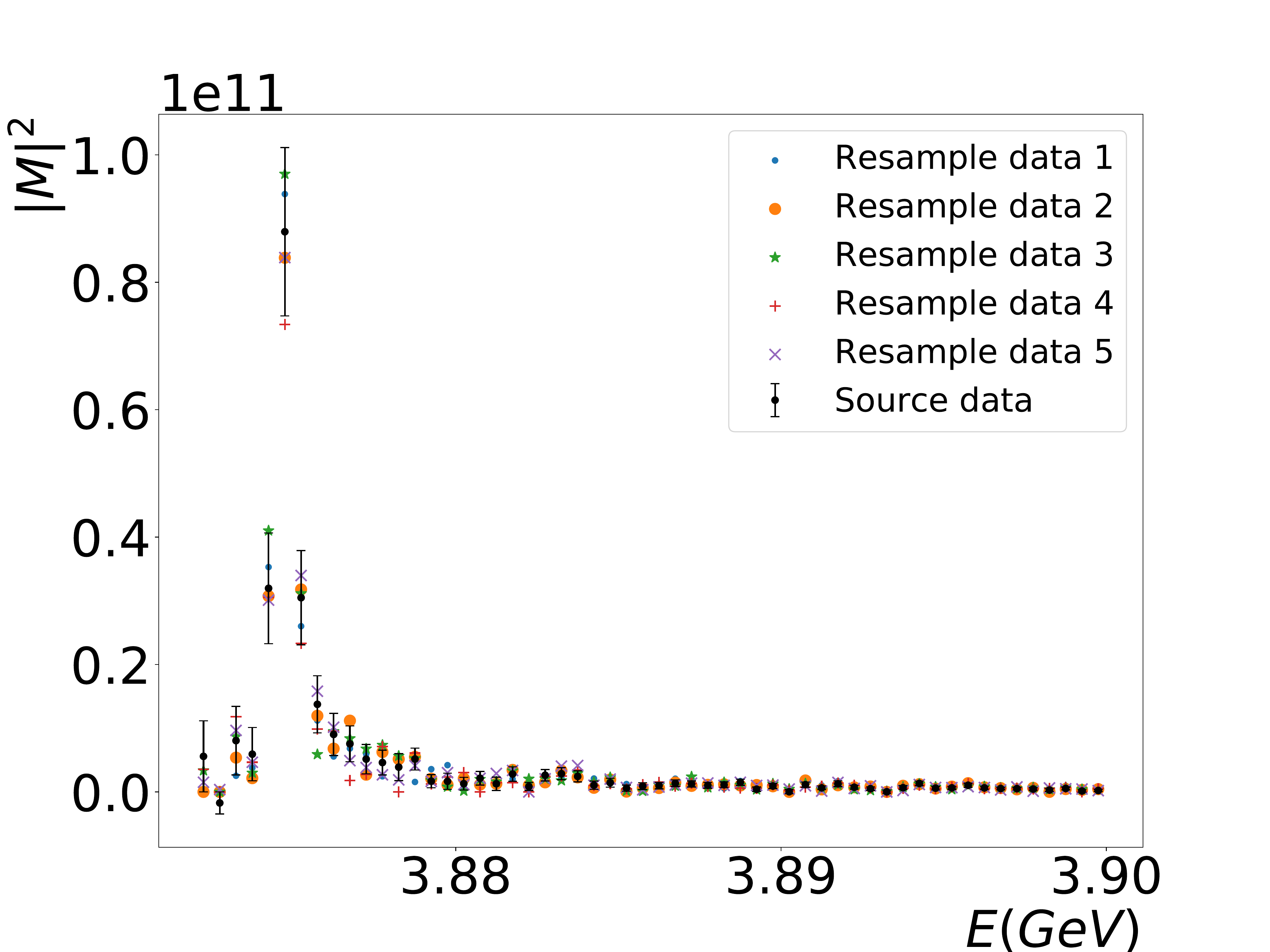}
    \caption{The resample data of the $X(3872)$ and the $T_{cc}^+$ comparing to the experimental data.
    The experimental data of the $X(3872)$ and the $T_{cc}^+$ are extracted from Ref.~\cite{Belle:2003nnu} and Refs.~\cite{LHCb:2021vvq,LHCb:2021auc}, respectively. }
    \label{X3872andTcc}
    \end{figure}

    \begin{figure}
\centering
\includegraphics[scale=0.35]{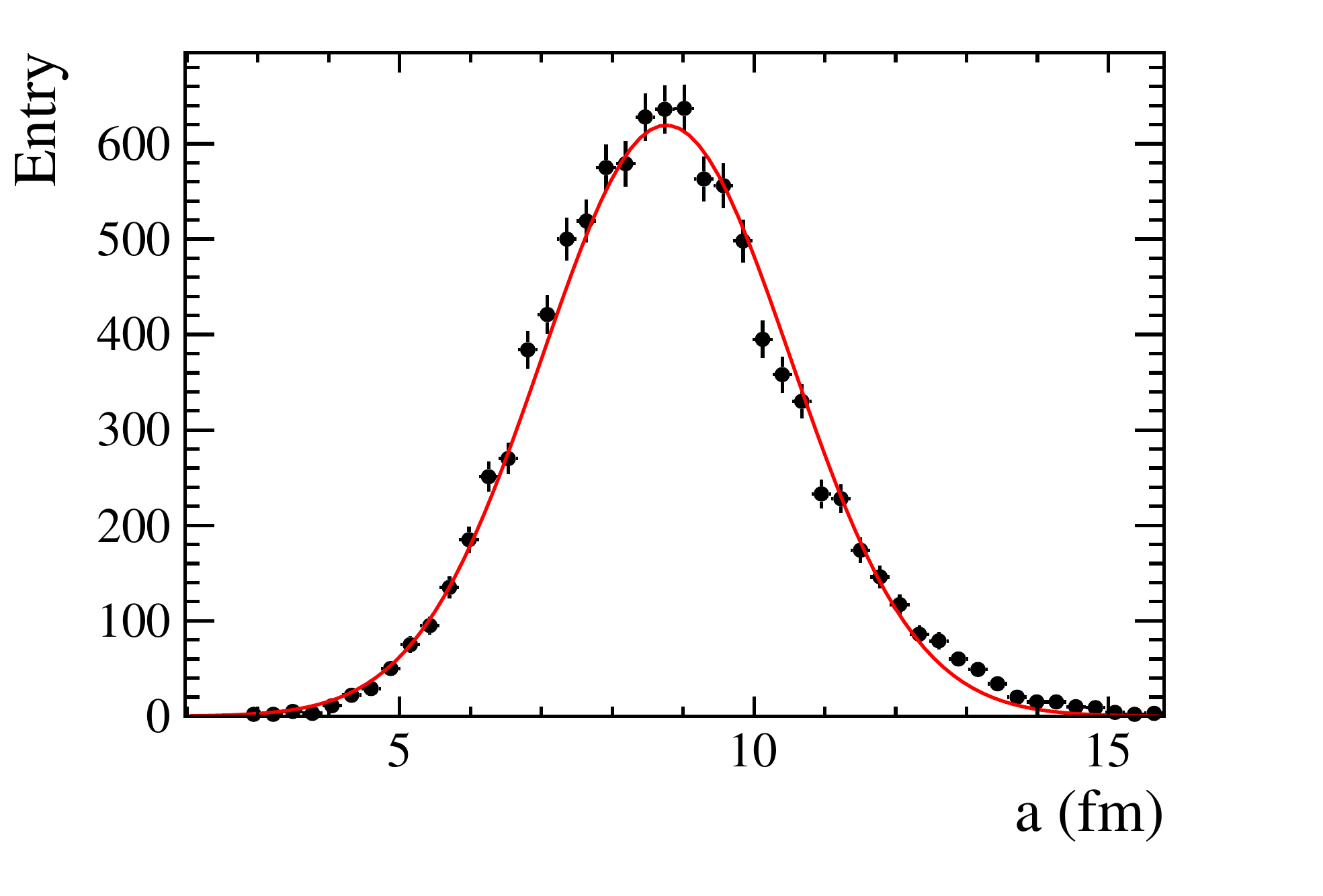}
\includegraphics[scale=0.35]{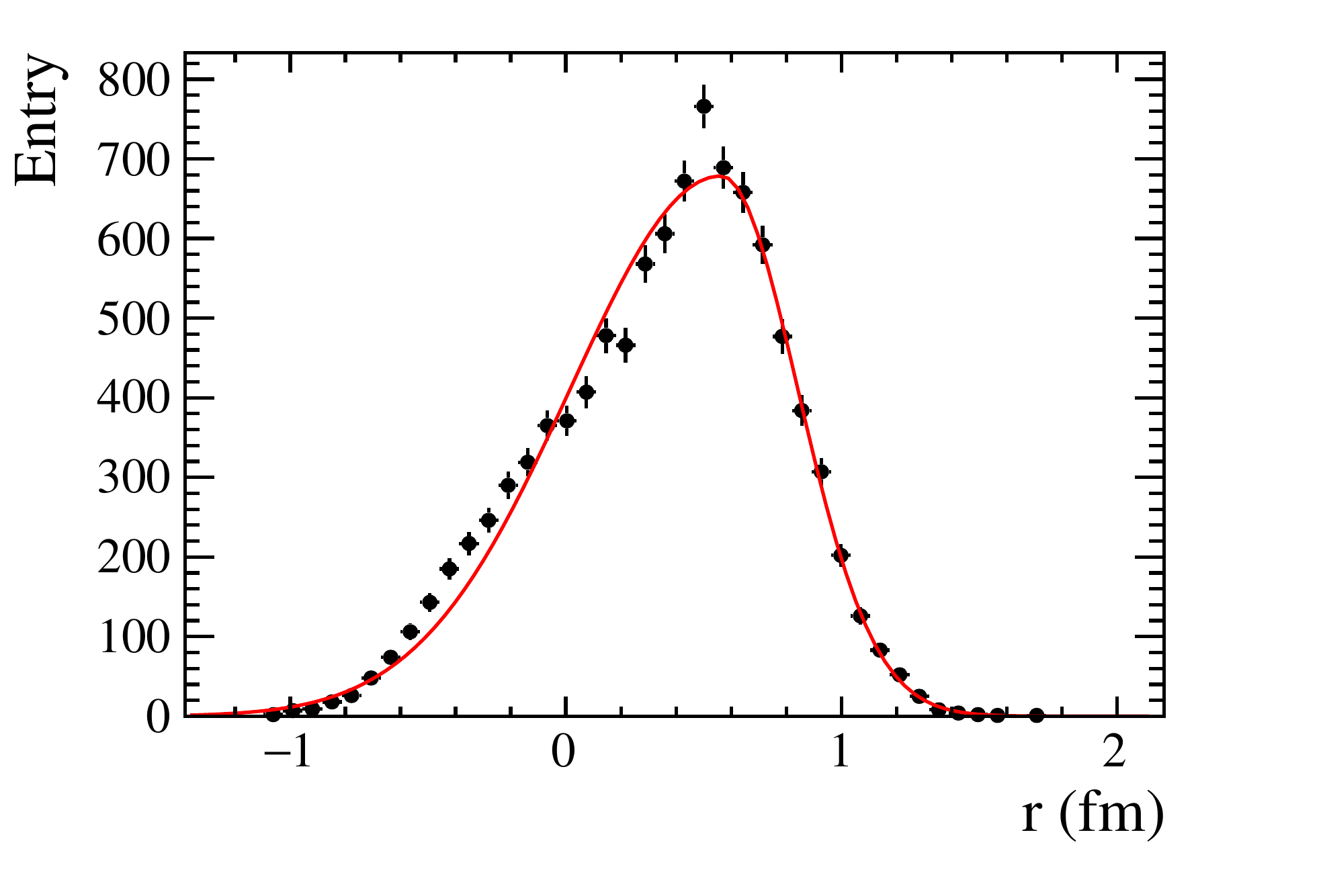}
\includegraphics[scale=0.35]{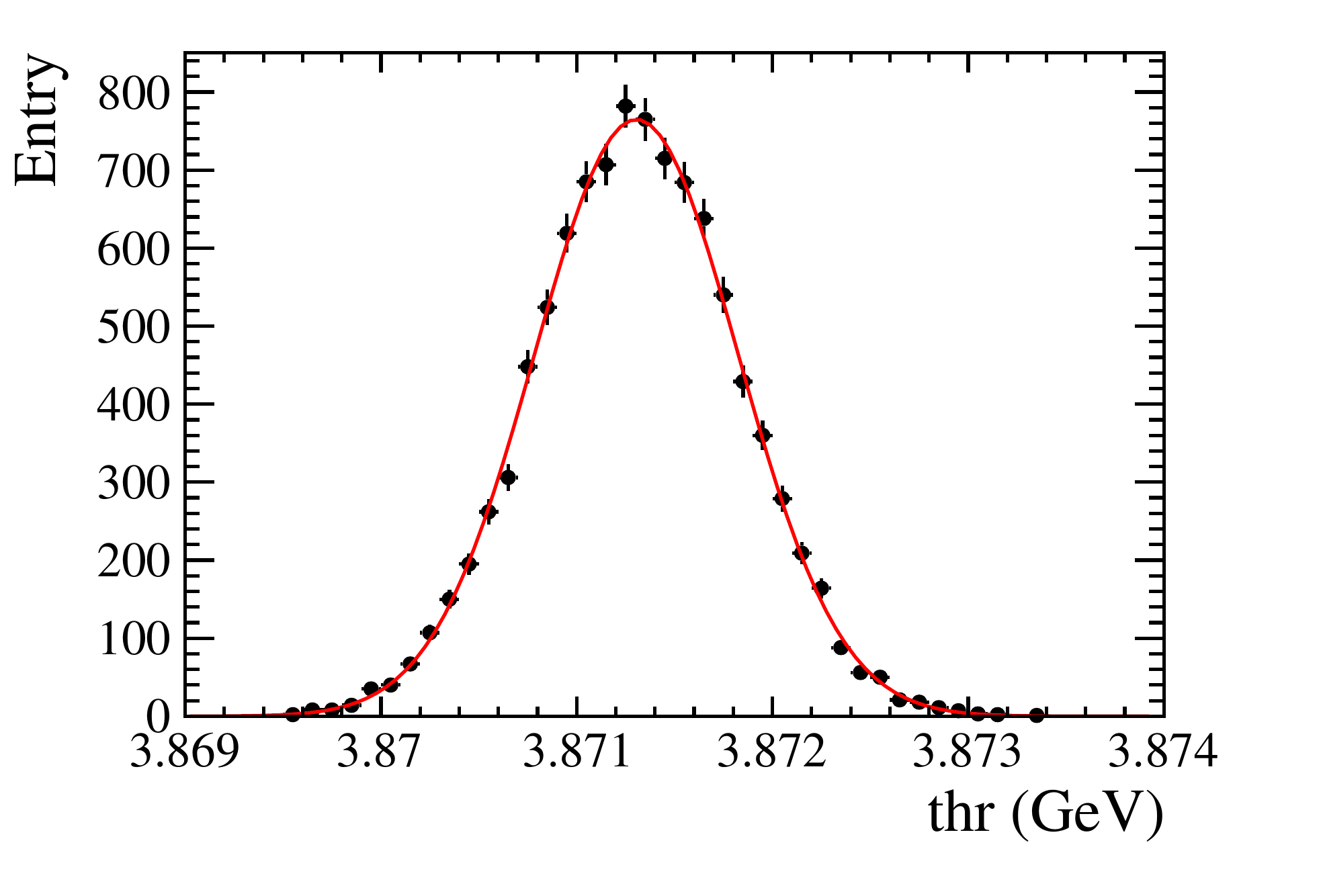}
\includegraphics[scale=0.35]{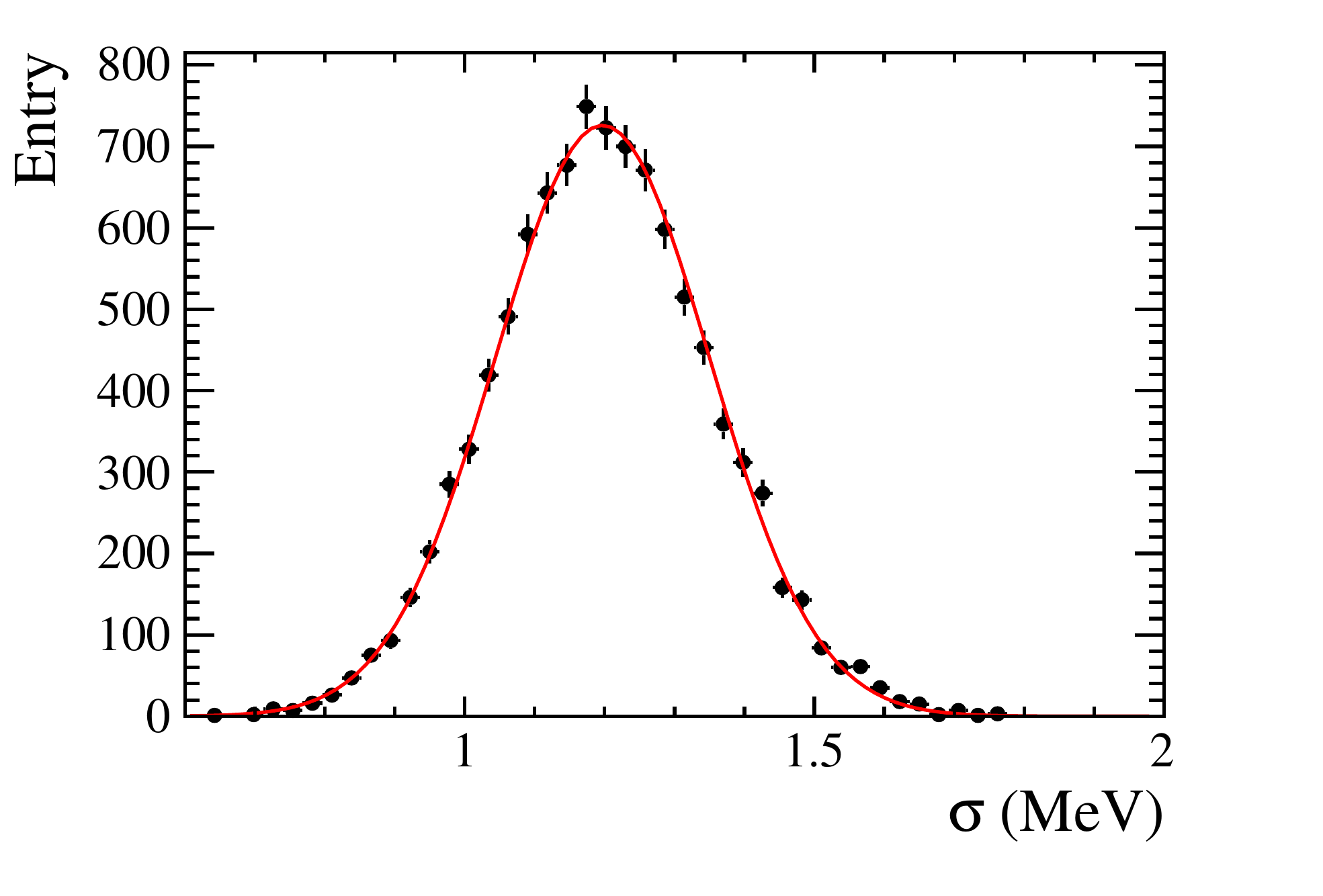}
\caption{Distributions of the predicted value of each parameter on $X(3872)$ experimental data. The red curves
are the fitted results with Gaussian function.}
\label{X3872par}
\end{figure}

\begin{figure}
\centering
\includegraphics[scale=0.35]{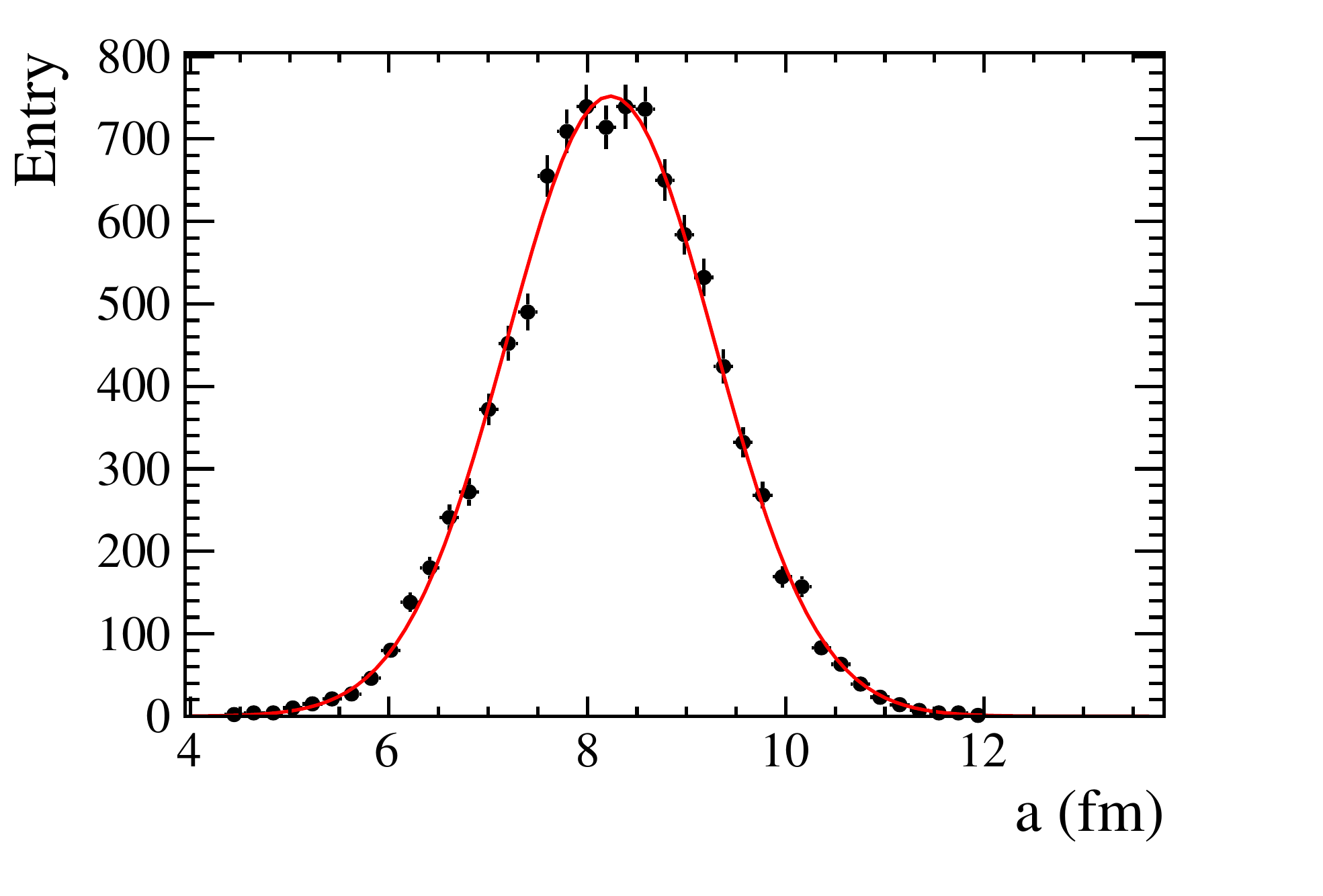}
\includegraphics[scale=0.35]{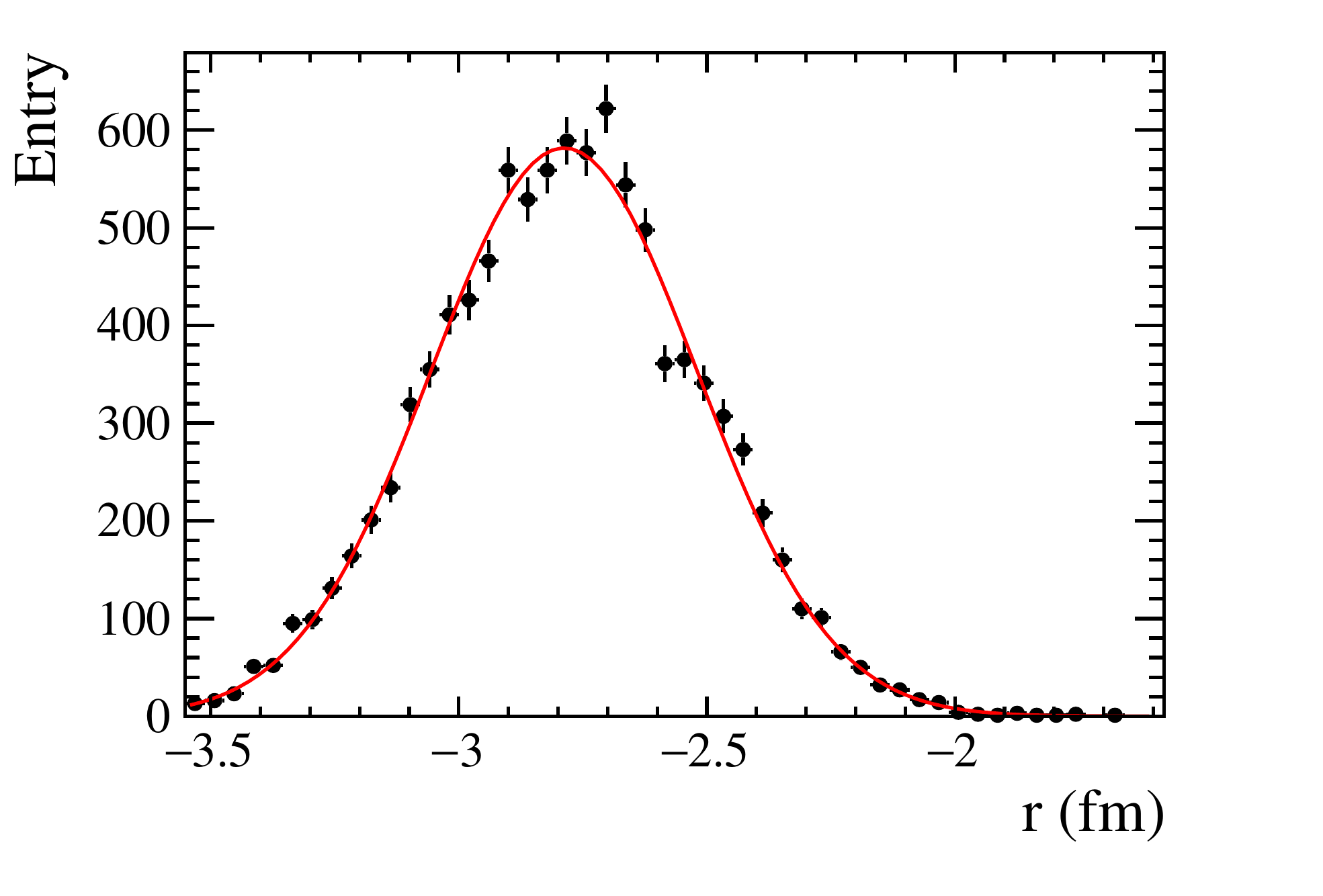}
\includegraphics[scale=0.35]{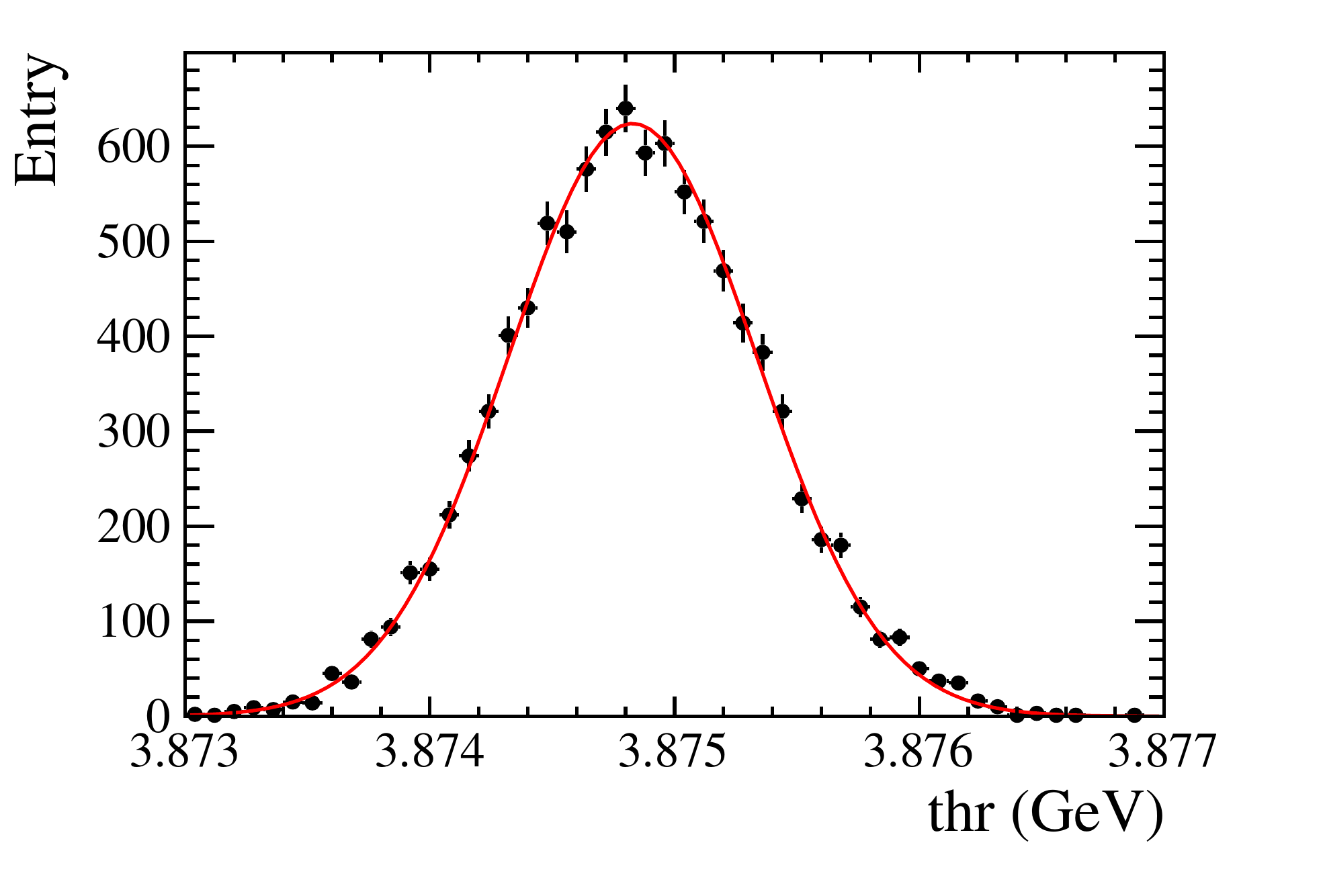}
\includegraphics[scale=0.35]{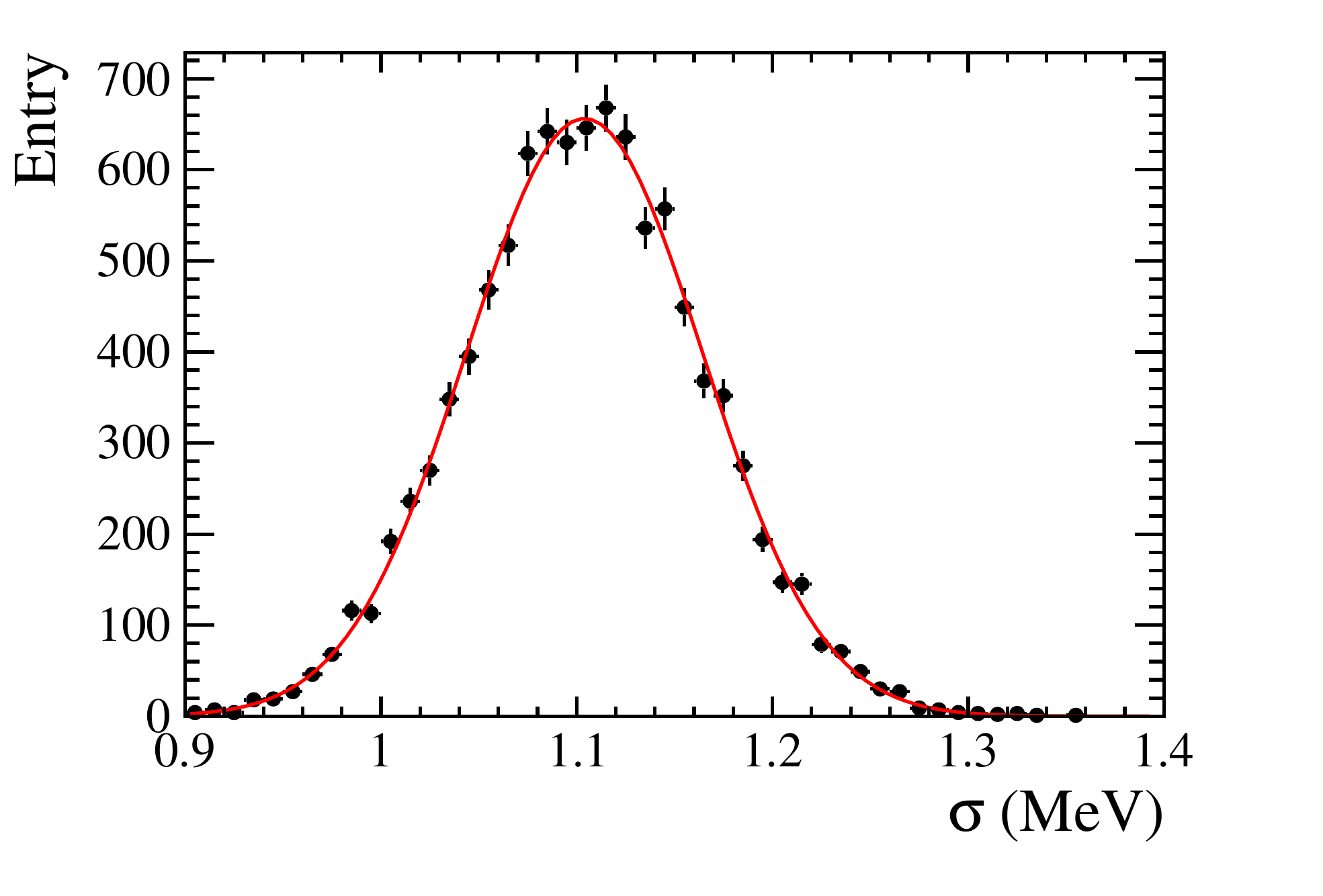}
\caption{Distributions of the predicted value of each parameter on $T_{cc}$ experimental data. The red curves
are the fitted results with Gaussian function.}
\label{Tccpar}
\end{figure}

Our network is applied to the experimental data of the
 $X(3872)$~\cite{Belle:2003nnu} and the $T_{cc}^+$~\cite{LHCb:2021vvq,LHCb:2021auc}
  with the three-body phase space subtracted,
 i.e., the $S$-wave $J/\psi\pi^+\pi^-$ and the $S$-wave $D^0D^0\pi^+$ phase space, respectively.
 The obtained parameters of the two states
 are collected in Tables~\ref{X3872 prediction} and Tables~\ref{Tcc prediction}, respectively,
 comparing to those from fitting with Eq.~\eqref{eq:ERE}.
 As shown in the two tables, the values of scattering lengths, effective ranges, and
 relevant thresholds from the two methods are consistent with each other
 within $1\sigma$ uncertainty. The resolution parameter $\sigma$ has
 a large deviation, which is because it is regressed individually and
 has larger uncertainty than those of scattering lengths and effective ranges.
  Especially, the absolute value of effective ranges of the $T_{cc}^+$ from the two methods
  are not as large as that in Refs.~\cite{LHCb:2021vvq,LHCb:2021auc}.
  The importance of this value is largely related to the nature of the $T_{cc}^+$~\cite{LHCb:2021auc,Baru:2021ldu,Du:2021zzh,Albaladejo:2021vln} .
  As a result, extracting this parameter precisely is valuable.
  Our network can also extract the most relevant threshold simultaneously.

The errors of the parameters are obtained in bootstrap~\cite{bootstrap}.
First, we resample the line shapes of the $X(3872)$ and the $T_{cc}^+$, as illustrated by Fig.~\ref{X3872andTcc}.
For the $i$th bin, a new event $Y_i = G(y_i, \sigma_i)$ is randomly generated
with the experimental central value $y_i$ and the experimental error $\sigma_i$
as  the mean and the standard deviation of the Gaussian probability density function (PDF).
Here, $G$ means sampling with the Gaussian PDF.
Here, 10000 $Y_i$ samples are generated and fed
into our deep learning approach. Five samples are plotted in Fig.~\ref{X3872andTcc}
comparing to the experimental data.
For a given parameter, all the predicted values form a Gaussian-like distribution,
as shown by Figs.~\ref{X3872par} and~\ref{Tccpar}.
The RMS of this distribution is taken as a quoted uncertainty,
  which is propagated from the experimental errors.

\section{Conclusion}
We train a neural network
to analyze the experimental mass spectra of exotic states.
The 150000 data samples are generated based on effective range expansion
and used for training the network, which can extract
scattering length, effective range, the most relevant threshold
, and the experimental resolution. The obtained parameters are consistent
with those from the fitting. The advantage of the neural network is that it is more stable than the fitting,
especially for low-statistic data. In addition, the compatibility of the neural network is
larger than the fitting. In principle, all theoretical models
can be encoded in one neural network, leaving it easier for experimental analysis.
As an application,
 the mass spectrum of the $X(3872)$ and the $T_{cc}^+$
 are studied. This network can also be applied for other one-channel near-threshold exotic states.

 \vspace{1cm}

\section{ACKNOWLEDGEMENT}
~~~This work is partly supported by Guangdong Major Project of Basic and Applied Basic Research Grant No.~2020B0301030008,
the National Natural Science Foundation of China Grant No.~12035007,
and Guangdong Provincial Grant No.~2019QN01X172.
Q.W. is also supported by the NSFC and the Deutsche Forschungsgemeinschaft (DFG, German
Research Foundation) through the funds provided to the Sino-German Collaborative
Research Center TRR110 ``Symmetries and the Emergence of Structure in QCD"
(NSFC Grant No. 12070131001, DFG Project-ID 196253076-TRR 110).

\end{document}